%% file: ulx_ism_revised.tex
\documentclass[fleqn,usenatbib]{mnras}

\usepackage{mathptmx}
\usepackage{txfonts}

\usepackage[T1]{fontenc}
\usepackage{ae,aecompl}

\usepackage{graphicx}	

\usepackage{float}

\newcommand{\beq}{\begin{equation}}
\newcommand{\eeq}{\end{equation}}
\newcommand{\beqa}{\begin{eqnarray}}
\newcommand{\eeqa}{\end{eqnarray}}

\newcommand{\NH}{N_{\rm{H}}}
\newcommand{\nH}{n_{\rm{H}}}

\newcommand{\Msun}{M_\odot}
\newcommand{\Rvir}{R_{\rm vir}}
\newcommand{\Tvir}{T_{\rm vir}}
\newcommand{\Vc}{V_{\rm c}}

\newcommand{\Mg}{M_{\rm g}}
\newcommand{\fg}{f_{\rm g}}
\newcommand{\rhog}{\rho_{\rm g}}
\newcommand{\rhogc}{\rho_{{\rm g}0}}
\newcommand{\rhomean}{\bar{\rho}}
\newcommand{\ve}{v_{\rm e}}

\newcommand{\mpr}{m_{\rm p}}
\newcommand{\kb}{k_{\rm B}}
\newcommand{\Omegab}{\Omega_{\rm b}}
\newcommand{\Omegam}{\Omega_{\rm m}}
\newcommand{\Omegal}{\Omega_\Lambda}

\newcommand{\tx}{t_{\rm X}}
\newcommand{\Lx}{L_{\rm{X}}}

\newcommand{\fx}{f_{\rm{X}}}

\newcommand{\Macc}{M_{\rm{acc}}}

\newcommand{\Ndot}{\dot{N}_\gamma}
\newcommand{\Nsmean}{\langle N_{\rm s}\rangle}
\newcommand{\RHII}{R_{\rm{HII}}}
\newcommand{\RHIImax}{R_{\rm{HII,max}}}
\newcommand{\alphaH}{\alpha_{\rm HII}}
\newcommand{\tHII}{t_{\rm HII}}

\newcommand{\nHe}{n_{\rm{He}}}
\newcommand{\RHeII}{R_{\rm{HeII}}}
\newcommand{\RHeIImax}{R_{\rm{HeII,max}}}
\newcommand{\alphaHeII}{\alpha_{\rm HeII}}
\newcommand{\tHeII}{t_{\rm HeII}}

\newcommand{\RHeIII}{R_{\rm{HeIII}}}
\newcommand{\RHeIIImax}{R_{\rm{HeIII,max}}}
\newcommand{\alphaHeIII}{\alpha_{\rm HeIII}}
\newcommand{\tHeIII}{t_{\rm HeIII}}

\newcommand{\xion}{x_{\rm{i}}}

\newcommand{\Ls}{L_{\rm{SX}}}
\newcommand{\Lsesc}{L_{\rm{SX,esc}}}
\newcommand{\Rs}{R_{\rm{SX}}}

\newcommand{\Lss}{L_{\rm{SSX}}}
\newcommand{\Lssesc}{L_{\rm{SSX,esc}}}
\newcommand{\Rss}{R_{\rm{SSX}}}

\newcommand{\Twind}{T_{\rm{wind}}}
\newcommand{\tsb}{t_{\rm sf}}

\title[Ultraluminous X-ray sources in the first
  galaxies]{Impact of ultraluminous X-ray sources on photoabsorption
  in the first galaxies}
\author[S. Sazonov and
  I. Khabibullin]{S. Sazonov$^{1,2}$\thanks{E-mail:
    sazonov@iki.rssi.ru} and I. Khabibullin$^{3,1}$\\ 
$^{1}$Space Research Institute, Russian Academy of Sciences,
  Profsoyuznaya 84/32, 117997 Moscow, Russia\\  
$^{2}$Moscow Institute of Physics and Technology, Institutsky per. 9, 
  141700 Dolgoprudny, Russia \\ 
$^{3}$Max Planck Institute for Astrophysics,
  Karl-Schwarzschild-Strasse 1, D-85741 Garching, Germany 
}

\begin{document}
\label{firstpage}
\pagerange{\pageref{firstpage}--\pageref{lastpage}}
\maketitle

\begin{abstract}
In the local Universe, integrated X-ray emission from high-mass X-ray
binaries (HMXBs) is dominated by the brightest ultraluminous X-ray sources
(ULXs) with luminosity $\gtrsim 10^{40}$~erg~s$^{-1}$. Such rare
objects probably also dominated the production of X-rays in the early
Universe. We demonstrate that a ULX with $\Lx\sim
10^{40}$--$10^{41}$~erg~s$^{-1}$ (isotropic-equivalent luminosity in
the 0.1--10~keV energy band) shining for $\sim
10^5$~years (the expected duration of a supercritically accreting
phase in HMXBs) can significantly ionise the ISM in its host dwarf
galaxy of total mass $M\sim 10^7$--$10^8\Msun$ and thereby reduce its
opacity to soft X-rays. As a result, the fraction of the soft X-ray
(below 1~keV) radiation from the ULX escaping into the intergalactic
medium (IGM) can increase from $\sim 20$--50\% to $\sim 30$--80\% over
its lifetime. This implies that HMXBs can induce a stronger heating of
the IGM at $z\gtrsim 10$ compared to estimates neglecting the ULX
feedback on the ISM. However, larger galaxies with $M\gtrsim
3\times 10^8\Msun$ could not be significantly ionised even by the
brightest ULXs in the early Universe. Since such galaxies probably
started to dominate the global star-formation rate at $z\lesssim 10$,
the overall escape fraction of soft X-rays from the HMXB population
probably remained low, $\lesssim 30$\%, at these epochs. 
\end{abstract}

\begin{keywords}
  stars: black holes -- accretion, accretion discs -- X-rays: binaries
  -- X-rays: ISM --  galaxies: high-redshift -- dark
  ages, reionization, first stars
\end{keywords} 

\section{Introduction}
\label{s:intro}

Already by $z\sim 10$, i.e. significantly before the Universe was
completely reionised (by $z\sim 6$), the intergalactic medium (IGM)
may have been heated above the temperature of the cosmic microwave
background (CMB), $T_{\rm CMB}=2.726 (1+z)$~K, by photoionising
radiation from the first generations of X-ray sources
(e.g. \citealt{venetal01,madetal04,ricost04,miretal11})\footnote{as
  well as by low-energy cosmic rays from the first supernovae
  \citep{sazsun15,leietal17} and microquasars
  \citep{tueetal14,douetal17}}. High-mass X-ray binaries (HMXBs) 
probably played the main role in this process, and many authors have
estimated the expected IGM temperature increment and
the associated 21~cm signal from the early Universe (most recently,
\citealt{madfra17,sazkha17a,rosetal17,cohetal17}). These  
predictions are important for preparing to 21~cm  observations of the 
reionisation epoch by upcoming experiments such as the Square
Kilometer Array (SKA\footnote{https://www.skatelescope.org}). 

Apart from the fairly uncertain specific (i.e. per unit star formation
rate, SFR) X-ray emissivity of HMXBs in the first galaxies, 
another poorly known aspect of the problem is whether the soft X-rays
emitted by HMXBs were actually able to escape into the IGM through the
interstellar medium (ISM) of their host galaxies. This is important
because only X-ray photons with energies $E\lesssim 1$~keV could
efficiently heat the Universe at $z\lesssim 20$, as the IGM was
essentialy transparent to harder X-ray radiation. Recently,
\citet{dasetal17} used high-resolution simulations of dwarf galaxies
at $z>7$ \citep{wisetal14}, taking into account radiative and 
supernova feedback from massive metal-free (Population~III) and metal
enriched stars on the ISM, to estimate neutral hydrogen column
densities through the host galaxies of HMXBs. These proved to vary
widely (by 1--2 orders of magnitude) from one line of sight to another and
between different galaxies, with median values of $\NH\sim 3\times
10^{21}$~cm$^{-2}$. Such column densities (in the essentially
metal-free medium of the first galaxies) imply that the soft X-ray
emission from HMXBs should be attenuated by photoabsorption in their
host galaxies by more than a factor of 3 at $E<500$~eV, significantly
weakening the X-ray heating of the IGM in the early Universe. Note
that \citet{dasetal17} assumed all helium (and heavier elements) in
the ISM to be neutral, while helium is the main photoabsorbing agent
of soft X-rays in a metal-poor medium. 

Virtually all previous studies were based on the assumption that 
HXMB X-ray emissivity is proportional to the SFR. Although this
conjecture is certainly reasonable when averaging over a large
ensemble of galaxies, it fails when considering galaxies
individually. Statistical studies of HMXBs and ultraluminous X-ray
sources (ULXs, usually defined as point-like, off-nucleus sources with
luminosity $\gtrsim 10^{39}$~erg~s$^{-1}$) in large
samples of nearby galaxies have demonstrated that the summed X-ray
emission of HMXBs and ULXs in the local Universe is dominated by the
brightest sources with luminosity $\gtrsim 10^{40}$~erg~s$^{-1}$
\citep{minetal12} or perhaps even $\sim 10^{41}$~erg~s$^{-1}$
(sometimes referred to as hyperluminous X-ray sources, HLXs), as the
measured {\sl intrinsic} HMXB/ULX X-ray luminosity function (LF)
continues with a slope $\alpha\sim 0.6$
($dN/d\log L\propto L^{-\alpha}$) up to $\sim 10^{40.5}$~erg~s$^{-1}$
(in the 0.25--8~keV energy band, and the same slope pertains to the LF
measured in the softer band of 0.25--2~keV), with no reliably detected cutoff
\citep{sazkha17b}. Recent studies indicate that the majority of ULXs
are an extension of the HMXB population towards higher, supercritical
accretion rates onto neutron stars and stellar-mass black holes (see
\citealt{kaaetal17} for a recent review), rather than accreting
intermediate-mass black holes. 

Although ULXs dominate the total X-ray output of HMXBs, such objects
are rare. Specifically, there are just $\sim 1.45$ ($\sim 0.36$)
sources with luminosity (absorption corrected, 0.25--8~keV) more than
$10^{39}$ ($10^{40}$)~erg~s$^{-1}$ per a SFR of 
1~$\Msun$~yr$^{-1}$ in the local Universe \citep{sazkha17b}, so that
many Milky Way sized galaxies do not contain such sources at all
(including the Milky Way itself, although the supercritical accretor
SS~433 would probably look like a ULX if we were viewing our Galaxy face-on,
\citealt{mesfab01,begetal06,pouetal07}; see, however, \citealt{khasaz16} for
upper limits on the collimated X-ray emission of SS~433). Assuming
that the shape of the HMXB/ULX LF in the early Universe was similar to
that at $z=0$ and given that the bulk of star formation at $z\sim 10$ was taking
place in dwarf galaxies with typical SFR $\lesssim
10^{-2}$~$\Msun$~yr$^{-1}$ (e.g. \citealt{xuetal16}), ULXs were even
rarer objects at those epochs: sources with luminosity $\gtrsim
10^{40}$~erg~s$^{-1}$ were present (at any given time) only in every
$\sim 100$th galaxy (even taking into account that the specific HMXB
occurrence rate was probably higher by a factor of $\lesssim 10$ due
the low metallicity of the first galaxies
(e.g. \citealt{broetal14,douetal15,baszyc16,lehetal16}). Nevertheless,
the X-ray background in the early Universe was probably dominated by
such extremely bright and rare 'lighthouses'!

Population synthesis modelling of massive binaries suggests that
supercritical accretion episodes with the X-ray luminosity reaching
$\sim 10^{40}$--$10^{41}$~erg~s$^{-1}$ can last up to a few~$\times
10^5$~yr \citep{rapetal05,pavetal17,wiketal17}. If such a ULX shined
in a dwarf galaxy, it might significantly photoionise the 
ambient ISM and thereby reduce the obscuration of its soft X-ray
radiation for an extra-galactic observer [there is some similarity
  here with supersoft X-ray sources (nuclear-burning white dwarfs)
photoionising the circumstellar matterial around them,
\citealt{niegil15}]. In particular, since in a neutral metal-poor
medium, X-rays are mostly absorbed by helium atoms, it is possible
that a bright ULX will primarily ionise helium rather than hydrogen in
the ISM, but that will still substantially weaken the attenuation of the ULX
radiation. Eventually, this would cause stronger IGM heating in the early
Universe. The goal of the present, proof-of-concept study is to
evaluate such ULX radiative feedback on the ISM of the first galaxies.

The following values of cosmological parameters are used below:
$\Omegam=0.309$, $\Omegal=1-\Omegam$, $\Omegab=0.049$,
$H_0=68$~km~s$^{-1}$~Mpc$^{-1}$ and $Y=0.246$ (helium mass fraction).

\section{Basic assumptions}
\label{s:assumptions}

The bulk of star formation in the cosmic X-ray heating epoch under
consideration ($z\sim 12$--8) was taking place in minihaloes and
haloes with total masses $\sim 10^7$--$10^9$~$\Msun$, with the lower
and higher values in this range being typical of the earlier and later
epochs, respectively (e.g. \citealt{xuetal16}). Below we consider a
situation where a single ULX appears in a star-forming halo and study
its photoionisation effect on the gas within the halo. 

\subsection{Properties of the halo and its gas}
\label{s:gas}

We consider a spherical dark matter halo of mass $M$ virialising at
redshift $z$, and assume that its density follows a
Navarro-Frenk-White (NFW) radial profile \citep{navetal97}:
\beq
\rho(r)=\frac{18\pi^2c^2\rhomean(z)}{3F(c)x(1+cx)^2}.
\label{eq:rho}
\eeq
Here, $\rhomean(z)$ is the mean matter density of the Universe at
redshift $z$, $x=r/\Rvir$, $\Rvir$ is the halo virial radius, $c$ is
the concentration parameter and $F(c)=\ln(1+c)-c/(1+c)$. We define
$\Rvir$ as the radius within which the average density is
$18\pi^2\approx 178$ times the critical density at the given
epoch. For the adopted cosmological parameters,
\beq
\Rvir=1.5\left(\frac{M}{10^8\Msun}\right)^{1/3}\left(\frac{1+z}{10}\right)^{-1}\,{\rm kpc}
\label{eq:rvir}
\eeq   
\citep{loefur13}.

We adopt $c=2$, since this value of the concentration parameter is
expected to be typical in the considered range of (high) redshifts and
(small) halo masses \citep{piletal17}. Following a number of previous 
theoretical studies of dwarf galaxies at the beginning of cosmic 
reionisation (e.g. \citealt{xuetal11,ferloe13}), we assume that the
gas in the halo has initial (prior to X-ray irradiation) temperature
$T_0$ and is in hydrostatic equilibrium within the dark matter
potential well 
\citep{maketal98}: 
\beq
\rhog=\rhogc e^{-(\Tvir/T_0)[\ve^2(0)-\ve^2(r)]/\Vc^2}.
\label{eq:rhog}
\eeq
Here, $\rhogc$ is the central gas density, $\Vc=(GM/\Rvir)^{1/2}$ is the
circular velocity at the virial radius, $\Tvir=(\mu\mpr/2\kb)\Vc^2$ is 
the virial temperature of the halo (where $\mu$ is the mean molecular
weight of the gas, $\mpr$ is the proton mass and $\kb$ is the
Boltzmann constant) and $\ve(r)$ is the escape velocity at radius $r$,
given by 
\beq
\ve^2(r)=2\Vc^2\frac{F(cx)+cx/(1+cx)}{xF(c)}.
\label{eq:ve}
\eeq
For the adopted cosmological parameters, the virial temperature is
given by \citep{loefur13}
\beq
\Tvir=1.0\times
10^4\frac{\mu}{0.6}\left(\frac{M}{10^8\Msun}\right)^{2/3}\frac{1+z}{10}\,{\rm K}.
\label{eq:tvir}
\eeq
Note that it can vary by a factor of 2 depending on the ionisation
state of the gas ($\mu\approx 0.6$ and 1.2 for a fully ionised and
neutral gas, respectively).

Equation~(\ref{eq:rhog}) allows for the possibility that the gas
temperature is different from the halo virial temperature. This factor
is aimed to crudely reflect the current uncertainty in the morphology
of the gas in high-redshift haloes associated with a multitude of
physical processes affecting their assembly, such as gas cooling,
large-scale motions and turbulence arising during halo virialisation
and mergers, angular momentum of the inflowing gas, stellar and
supernova feedback etc. Cosmological hydrodynamical simulations suggest that
high-redshift dwarf galaxies and minihaloes were quite irregular but
their overall morphology was rounder compared to the more massive
galaxies at later epochs and most of them probably did not form
well-defined central discs
(e.g. \citealt{wisabe07,greetal08,wiscen09,rometal11,wisetal14}). 

From the above formulae, the gas density can finally be expressed as
\beq
\rhog(x)=\rhogc e^{-A \Tvir/T_0}(1+cx)^{A(\Tvir/T_0)/cx},
\label{eq:rhogf}
\eeq
where $A=2c/F(c)$. The central gas density is determined by equating
the total gas fraction within the virial radius, $\fg\equiv\Mg/M$, to
the cosmological value, $\Omegab/\Omegam\approx 0.16$; thus 
\beq
\rhogc=\frac{(18\pi^2/3)\fg c^3e^{A\Tvir/T_0}}{\int_0^c
  t^2(1+t)^{A(\Tvir/T_0)/t}\,dt}\rhomean(z).
\label{eq:rhog0}
\eeq

Since high-redshift dwarf galaxies and minihaloes were very metal
poor, we assume, in our baseline model, the gas to consist of hydrogen
and helium in their primordial mass ratio (0.754 to 0.246). However,
we have also carried out tests to check the sensitivity of our results
to the presence of small amounts of metals in the gas. 

We do not consider the gas outside the halo,
since its X-ray absorption optical depth is expected to be
small. Indeed, since the average hydrogen number density in the
Universe $\langle{\nH}\rangle\sim 3\times
10^{-4}[(1+z)/10]^3$~cm$^{-3}$ and the density outside the halo
gradually declines from $\sim 4\langle\nH\rangle$ at $\Rvir$ to less
than $2\langle\nH\rangle$ at a few $\Rvir$
(e.g. \citealt{bertschinger85,medetal16}), the total column density
through the IGM outside the halo is $\lesssim {\rm
  a~few}~10^{19}(M/10^8\Msun)^{1/3}[(1+z)/10]^2$~cm$^{-2}$, much less
than $\NH\sim 10^{21}$--$10^{22}$~cm$^{-2}$ expected for the halo
itself. 

\subsection{ULX properties}
\label{s:ulx}

Suppose that a ULX switches on at the centre of the halo (during a
cosmologically short star-formation episode) and emits an X-ray
luminosity $\Lx$ in the energy range from 100~eV to 10~keV 
for a time $\tx$. We assume that the source's X-ray spectrum has a
power-law shape with a photon index $\Gamma=2.1$ 
($dN_\gamma/dE\propto E^{-\Gamma}$). Such a spectrum fits well the
absorption-corrected summed X-ray spectrum of luminous ($\Lx\gtrsim
10^{38}$~erg~s$^{-1}$, including ULXs) HMXBs in the local Universe and
represents the angle-integrated X-ray emission of the near- and
supercritically accreting neutron stars and stellar-mass black holes 
at $z=0$ \citep{sazkha17c}.

This spectrum was originally obtained from {\sl Chandra} X-ray
Observatory data in the 0.25--8~keV band. Here, we extrapolate it to a
somewhat broader energy range, in particular to lower energies,
because (i) theoretical models of supercritical accretion
(e.g. \citealt{watetal00,pouetal07,kawetal12,vinetal13,naretal17}),
(ii) observations of extended (several hundred parsecs) high-ionisation
nebulae around ULXs
(e.g. \citealt{pakmir03,kaaetal04,beretal10})\footnote{Some of these
  nebulae may be close analogs of the ionisation bubbles inflated by
  ULXs in the first galaxies, discussed in the present study.} and
(iii) optical/UV observations of the Galactic microquasar SS~433
\citep{cheetal82,doletal97,fabrika04} suggest that the supercritical
disc with an outflowing wind emits at soft X-ray/UV energies a
luminosity comparable to the X-ray luminosity emergent from the
central disc funnel.
Our choice of the value of 100~eV for the lower boundary of the
  spectral energy range is rather arbitrary and partly driven by the
  fact that the apparently power-law shape of the aforementioned
  collective X-ray spectrum of luminous HMXBs \citep{sazkha17c}
  results from averaging over very diverse spectra (from hard to
  supersoft) of individual sources, and it is not obvious how to
  extrapolate such a summed spectrum substantially below its original lower
  energy boundary (250~eV). In addition, this choice is motivated by
  the estimated temperature of the accretion disc's wind
  in SS~433 of $\Twind\sim 5\times 10^4$~K and the expectation (based
  on a theoretical scaling relation of $\Twind$ with the supercritical
  accretion rate) that the winds in typical ULXs may be hotter by a
  factor of a few, i.e. $\Twind\lesssim 2\times 10^5$~K
  \citep{fabetal15}. One may therefore expect a substantial
  fraction of the X-ray radiation emitted by the supercritical
  accretion disk to be reprocessed by the wind at energies $E\sim
  4k\Twind\lesssim 100$~eV.
  
Importantly, the quantity $\Lx$ introduced above is not meant to be
the angle-integrated luminosity of the ULX. In fact, the ULX 
radiation is likely to be collimated along the axis of the supercritical
accretion disc (e.g. \citealt{king09}) and then by $\Lx$ one should
understand an 'isotropic-equivalent luminosity', i.e. $\Lx=4\pi 
D^2\fx$ (where $D$ is the distance to the source and $\fx$ is the
measured X-ray flux), which would be perceived by an observer viewing
the disc nearly face-on. It is this apparent luminosity that is
relevant for the present study, since our goal is to explore the
photoionisation effect of the ULX on the gas located in front of its
X-ray beam, which further irradiates the surrounding IGM. In reality,
the intensity and spectral shape of the radiation emitted by the ULX
is likely to vary with the offfset angle from the axis
of the disc in a more complicated way, but since this dependence is
poorly known and moreover can depend on the accretion rate and the
nature of the relativistic compact object (black hole vs. neutron
star), we ignore it in our modelling.

The key parameters of the model are thus $\Lx$ and $\tx$. To get an
idea of how large their values can be, suppose that a ULX is powered
by accretion from a stellar companion onto a relativistic compact
object, the radiative efficiency of accretion is $\eta$ and the X-ray
radiation is collimated within two opposite cones subtending a total
solid angle $\Omega$. Then, the total mass accreted from the companion
(part of which may be ejected as a wind from the system and not reach
the compact object) will be 
\beqa
\Macc &=& \frac{\Lx\tx}{\eta c^2}\frac{\Omega}{4\pi}\\
\nonumber
&\approx& 1.8\frac{\Lx}{10^{41}\,{\rm erg\,s}^{-1}}\frac{\tx}{10^5\,{\rm yr}}\frac{0.01}{\eta}\frac{\Omega/4\pi}{0.1}~\Msun.
\label{eq:macc}
\eeqa
By adopting $\eta=1$\% in the second part of this expression, we
crudely took into account the expected low radiative efficiency of
supercritical accretion
(e.g. \citealt{shasun73,jaretal80,watetal00,pouetal07}) compared  
to standard, subcritical accretion ($\sim
10$\%). Equation~(\ref{eq:macc}) implies that if a moderate,
$\Omega/4\pi\sim 0.1$, degree of X-ray collimation is achieved, high
luminosities $\Lx\lesssim 10^{41}$~erg~s$^{-1}$ (note that for a
power-law spectrum with $\Gamma=2.1$, $\Lx=10^{41}$~erg~s$^{-1}$ in
our adopted 0.1--10~keV energy band corresponds to just $\sim 3\times
10^{40}$~erg~s$^{-1}$ in the standard 2--10~keV X-ray band) can be
sustained for a long time $\tx\sim 10^5$~yr through the accretion of a
reasonable amount of matter (few $\Msun$) from the companion (massive)
star. These crude estimates are supported by detailed population
synthesis studies of massive stellar binaries \citep{pavetal17,wiketal17}.   

\section{Analytical estimates}
\label{s:estimates}

Before proceeding to detailed numerical calculations, it is useful to
obtain order-of-magnitude estimates for the ULX feedback on the ISM.

\subsection{Ionisation of hydrogen}
\label{s:HII}

Suppose first that the ISM consists purely of neutral hydrogen
with constant number density $\nH$ and temperature $T_0=10^4$~K. In this
case, the ULX, after switching on, will start inflating an HII
zone. The radius of the latter will grow with time approximately
(neglecting partial absorption of the ionising radiation within this
radius) according to the differential equation $\Ndot(1+\Nsmean)
\,dt=4\pi\nH\RHII^2\,d\RHII$, where $\Ndot$ is the apparent
(i.e. isotropic-equivalent) number of hydrogen-photoionising
($E>13.6$~eV) photons emitted by the ULX per second and $\Nsmean$ is
the average number of secondary ionisations induced by the primary
photoelectron. Hence,  
\beq
\RHII(t)=\left[\frac{3\Ndot(1+\Nsmean) t}{4\pi\nH}\right]^{1/3}.
\label{eq:rHII}
\eeq

The maximum possible size of such an HII zone, determined by the
balance  between ionisations and recombinations, is given by the
classical Str\"{o}mgren formula: 
\beq
\RHIImax=\left[\frac{3\Ndot}{4\pi\nH^2\alphaH}\right]^{1/3},
\label{eq:rHIImax}
\eeq
where $\alphaH\approx 2.54\times 10^{-13}$~cm$^3$~s$^{-1}$ is the Case
B recombination rate for hydrogen at $T=10^4$~K \citep{draine11}. Note
that we have omitted the $(1+\Nsmean)$ factor in this expression, in
contrast to equation~(\ref{eq:rHII}), since a stationary HII zone is
characterised by a high ionisation fraction (i.e. the fraction of free
electrons) $\xion\approx 1$, so that secondary ionisations play a
negligible role (the photoelectron deposits most of its energy into
heat).  

For our adopted X-ray spectrum (power-law with $\Gamma=2.1$ from
100~eV to 10~keV), all photons have energies above the hydrogen
ionisation threshold, so that $\Ndot\approx \Lx/370~{\rm   eV}\approx
1.7\times 10^{50}(\Lx/10^{41}\,{\rm erg~s}^{-1})$~s$^{-1}$ (here 370~eV
is the ULX spectrum-weighted average energy of ionising photons). In a
nearly neutral medium (with $\xion<0.01$, as e.g. expected for a
collisionally ionised ISM at $T\sim 10^4$~K), the fraction of the
energy of the primary photoelectron (the latter being $E-13.6~{\rm
  eV}\approx E$) deposited into secondary ionisations is approximately
constant, $\sim 0.3$, at energies above 100~eV \citep{fursto10,valfer08}, so
that $\Ndot(1+\Nsmean)\approx \Lx/370\,{\rm eV}+0.3\Lx/13.6\,{\rm
  eV}\approx (0.17+1.38)\times 10^{51}(\Lx/10^{41}\,{\rm
  erg~s}^{-1})$~s$^{-1} \approx 1.55\times 10^{51}(\Lx/10^{41}\,{\rm
  erg~s}^{-1})$~s$^{-1}$. Hence, 
\beq
\RHII(t)\approx 158\left(\frac{\Lx}{10^{41}~{\rm
    erg~s}^{-1}}\right)^{1/3}\left(\frac{t}{10^5~{\rm 
      yr}}\right)^{1/3}\left(\frac{\nH}{10~{\rm
      cm}^{-3}}\right)^{-1/3}~{\rm pc}
\label{eq:rHIIpc}
\eeq
and
\beq
\RHIImax\approx 38\left(\frac{\Lx}{10^{41}~{\rm
    erg~s}^{-1}}\right)^{1/3}\left(\frac{\nH}{10~{\rm
      cm}^{-3}}\right)^{-2/3}~{\rm pc}.
\label{eq:rHIImaxpc}
\eeq
Therefore, to inflate the maximum possible HII zone
($\RHII=\RHIImax$), the ULX must be active for a time longer than
\beq
\tHII\approx 1.4\times 10^3\left(\frac{\nH}{10~{\rm cm}^{-3}}\right)^{-1}~{\rm
  yr},
\label{eq:tHII}
\eeq
which does not depend on the ULX luminosity. 

\subsection{Ionisation of helium}
\label{s:He}

Despite the low abundance of helium compared to hydrogen in the ISM
(their primordial number ratio is $\sim 8$ to 92), a soft X-ray
($E\sim 100$--500~eV) photon emitted by the ULX is $\sim 2$ times more
likely to be absorbed by an HeI atom than by a HI atom. Therefore, the
ULX might inflate a larger HeII zone compared to a HII zone.  

Following the discussion for hydrogen above, the
time-dependent and maximum sizes of the HeII zone can be estimated as
\beq
\RHeII(t)=\left[\frac{3\Ndot(1+\Nsmean) t}{4\pi\nH}\frac{\nH}{\nHe}\right]^{1/3}
\label{eq:rHeII}
\eeq
and
\beq
\RHeIImax=\left[\frac{3\Ndot}{4\pi\nH^2\alphaHeII}\left(\frac{\nH}{\nHe}\right)^2\right]^{1/3},
\label{eq:rHeIImax}
\eeq
respectively, where $\alphaHeII\approx 2.72\times
10^{-13}$~cm$^3$~s$^{-1}$ is the case B recombination rate for 
helium at $T=10^4$~K \citep{draine11}.

Compared to the case of hydrogen ionisation, the $(1+\Nsmean)$ factor 
 plays a much less important role for helium, because the
majority of secondary ionisations induced by the fast electron
resulting from the photoionisation of a He atom will take place on
H rather than He atoms. Specifically, in nearly neutral medium
($\xion<0.01$), the fraction of the energy of the photoelectron
deposited into secondary ionisations of helium is $\sim 0.025$ for
energies $\sim 100$--500~eV \citep{fursto10}\footnote{Note that we
  have omitted the $(1+\Nsmean)$ 
  factor in equation~(\ref{eq:rHeIImax}) for $\RHeIImax$, since even
  if the hydrogen remains mostly neutral, the HeII zone will be
  characterised by $\xion>0.1$.}. Therefore, given the HeI ionisation
threshold of 24.6~eV, $\Ndot(1+\Nsmean)\approx \Lx/370\,{\rm
  eV}+0.025\Lx/24.6\,{\rm eV}\approx (1.7+0.6)\times 10^{50}(\Lx/10^{41}\,{\rm
  erg~s}^{-1})$~s$^{-1}\approx 2.3\times 10^{50}(\Lx/10^{41}\,{\rm
  erg~s}^{-1})$~s$^{-1}$, and thus 
\beq
\RHeII(t)\approx 189\left(\frac{\Lx}{10^{41}~{\rm
    erg~s}^{-1}}\right)^{1/3}\left(\frac{t}{10^5~{\rm
    yr}}\right)^{1/3}\left(\frac{\nH}{10~{\rm
    cm}^{-3}}\right)^{-1/3}~{\rm pc}; 
\label{eq:rHeIIpc}
\eeq
\beq
\RHeIImax\approx 189\left(\frac{\Lx}{10^{41}~{\rm
    erg~s}^{-1}}\right)^{1/3}\left(\frac{\nH}{10~{\rm
      cm}^{-3}}\right)^{-2/3}~{\rm pc}.
\label{eq:rHeIImaxpc}
\eeq
The maximum possible size of the HeII zone can be achieved in a time
\beq
\tHeII\approx 1.0\times 10^5\left(\frac{\nH}{10~{\rm cm}^{-3}}\right)^{-1}~{\rm
  yr}.
\label{eq:tHeII}
\eeq

Similar considerations can be applied to the case of a HeIII (fully
ionised helium) zone. In this case, one can safely neglect the
$(1+\Nsmean)$ factor, so that 
\beq
\RHeIII(t)\approx 171\left(\frac{\Lx}{10^{41}~{\rm
    erg~s}^{-1}}\right)^{1/3}\left(\frac{t}{10^5~{\rm
    yr}}\right)^{1/3}\left(\frac{\nH}{10~{\rm
    cm}^{-3}}\right)^{-1/3}~{\rm pc}; 
\label{eq:rHeIIIpc}
\eeq
\beq
\RHeIIImax\approx 94\left(\frac{\Lx}{10^{41}~{\rm
    erg~s}^{-1}}\right)^{1/3}\left(\frac{\nH}{10~{\rm
      cm}^{-3}}\right)^{-2/3}~{\rm pc};
\label{eq:rHeIIImaxpc}
\eeq
\beq 
\tHeIII\approx 1.7\times 10^4\left(\frac{\nH}{10~{\rm cm}^{-3}}\right)^{-1}~{\rm yr}.  
\label{eq:tHeIII}
\eeq 
Here, we have taken into account the (relatively high) HeIII
$\rightarrow$ HeII recombination rate $\alphaHeIII\approx 2.19\times
10^{-12}$~cm$^3$~s$^{-1}$ at $T=10^4$~K \citep{draine11}.

\subsection{Preliminary conclusions and notes of caution}
\label{s:preliminary}

We can therefore draw the following preliminary conclusions: (i) a ULX
can inflate a maximum-size ionisation zone within its lifetime
($\tx\sim 10^5$~yr) unless the typical ISM density $\nH\ll
10$~cm$^{-3}$, (ii) the HII-zone is expected to be somewhat 
smaller compared to the HeII and HeIII zones, and (iii) for the most
luminous ULXs ($\Lx\sim 10^{41}$~erg~s$^{-1}$), the HeII/HeIII bubble
can grow large enough ($\sim 100~{\rm pc}\sim 0.1\Rvir$) to
significantly [by $\Delta\NH\sim 3\times 10^{21}(\Lx/10^{41}\,{\rm
  erg~s}^{-1})^{1/3}(\nH/10\,{\rm cm}^{-3})^{1/3}$~cm$^{-2}$] reduce
the total X-ray absorption column density in front of the ULX,
especially in smaller haloes.

In reality, the ionisation of hydrogen and helium will proceed in a
joint fashion. Moreover, HeII $\rightarrow$ HeIII ionisation follows
HeI $\rightarrow$ HeII ionisation, rather than proceeds
simultaneously. Therefore, the   
estimates presented above in \S\ref{s:HII} and \S\ref{s:He} should be
taken with caution. Also, the above treatment in terms of
classical Str\"{o}mgren spheres may not be fully adequate for the
problem at hand, since for typical soft X-ray energies playing a role
here (50\% and 90\% of all photons have $E<200$~eV and $E<800$~eV,
respectively), the photoionisation cross-section in a neutral H+He
gas\footnote{This power-law dependence, derived from the more general
  and accurate formulae by \citealt{veretal96}, is a good approximation at
  $100<E\lesssim 1000$~eV.} $\approx 6\times 10^{-21}(E/200~{\rm
  eV})^{-3.2}$~cm$^{-2}$, so that the photon mean free path is
\beq
\lambda\approx 5\left(\frac{\nH}{10~{\rm cm}^{-3}}\right)^{-1}\left(\frac{E}{200~{\rm eV}}\right)^{3.2}~{\rm pc}.
\label{eq:lambda}
\eeq
Therefore, the condition $\lambda\ll (\RHIImax, \RHeIImax,
\RHeIIImax)$ will be fulfilled only if the gas is dense enough and/or
the ULX is sufficiently luminous, namely if $\nH\gg
0.1(\Lx/10^{41}\,{\rm erg~s}^{-1})^{-1}$~cm$^{-3}$. Otherwise, instead
of a well-defined separation between a central ionised zone and the
ambient nearly neutral gas, the ULX will cause a spatially distributed
partial ionisation of the ISM.

Moreover, our assumption of an initially nearly neutral ISM may not
be suitable for the host galaxies of ULXs in the early
Universe. Indeed, a substantial fraction of the ISM may have
temperatures significantly higher than $10^4$~K in relatively massive
($M\gtrsim 10^8\Msun$) haloes as a result of their virialisation (see
equation~\ref{eq:tvir}), in which case collisional ionisation will
lead to a non-negligible ionisation fraction $\xion$. In addition, the
ISM may be pre-ionised by UV radiation and shock waves associated with
star formation in the galaxy. All these processes, however, are
unlikely to lead to a significant ionisation of helium in the
ISM. Hence, our consideration of X-ray-driven helium ionisation should
remain practically unaltered in this case. 
 
\subsection{ISM heating}
\label{s:heat}

The ULX radiation will not only ionise but also heat the surrounding ISM. 

We may crudely estimate the amplitude of this effect using an
ionisation parameter, $U=\Ndot/4\pi R^2\nH c$\footnote{In the
  considered case, the lower boundary of the spectral range,
  100~eV, is higher than the ionisation threshold of
  hydrogen.}. Taking $R\sim 100$~pc -- the expected size of the
ionisation zone, and given that $\Ndot\approx 1.7\times
10^{50}(\Lx/10^{41}\,{\rm erg~s}^{-1})$~s$^{-1}$ for the adopted ULX
spectrum, we find that $U\sim 5\times 10^{-4} (\Lx/10^{41}\,{\rm
  erg~s}^{-1})(R/100\,{\rm pc})^{-2}(\nH/10\,{\rm cm}^{-3})^{-1}$. As is
well known \citep{draine11}, for $U\sim 10^{-5}$--$10^{-2}$,
the gas equilibrium temperature (determined by the balance between
photoionisation heating and various cooling processes) is $\sim
(1-2)\times 10^{4}$~K. We may thus expect that even if the
temperature of the ISM is initially lower than $10^4$~K, the ULX
will heat it to a~few~$10^{4}$~K at distances $\lesssim 100$~pc. 

We finally note that the sound-crossing time of the ionisation zone
(with $T\sim 10^4$~K) is expected to be $\sim 100~{\rm pc}/10~{\rm 
  km~s}^{-1}\sim 10^7$~yr, much longer than the ionisation time
(i.e. the ULX lifetime), $\tx\lesssim 10^5$~yr. One can therefore
consider the ionised gas to be stationary, despite it being
over-pressured relative to the surrounding gas.
    
\section{Numerical calculations}
\label{s:calculations}

To self-consistently compute the evolution of the expanding H/He
ionisation region and the attenuation of the ULX emission by gas
with the hydrostatic density distribution defined in \S\ref{s:gas}
above, we took advantage of the publicly available numerical package
{\sc Cloudy} (version 17.00), which makes it possible to perform
photoionisation and radiative transfer computations in a time
dependent manner \citep{feretal17}. 

We start with defining several galaxy models (see
Table~\ref{tab:galaxies}) that may be regarded as representative
examples in the context of the early X-ray heating  of the Universe. To
this end, we considered three reference epochs: $z=12.5$, 10 and
8. Here, the first value is 
approximately the redshift when HMXBs probably became numerous enough
to begin heating the IGM efficiently
(e.g. \citealt{madfra17,sazkha17a}), $z\approx 8$ approximately marks 
the beginning of intense re-ionization of the Universe by UV radiation
(so that X-ray heating becomes relatively unimportant), and $z=10$ may
be considered a typical redshift of the X-ray heating
epoch. Furthermore, this choice is convenient because the same
redshifts were adopted as reference ones in the recent simulations of
star formation in the early Universe by \cite{xuetal16}, which largely
governed our choice of fiducial galaxy masses below.

\input{galaxies_table.tex}

\begin{figure}
\centering
\includegraphics[width=\columnwidth,viewport=20 160 560
  700]{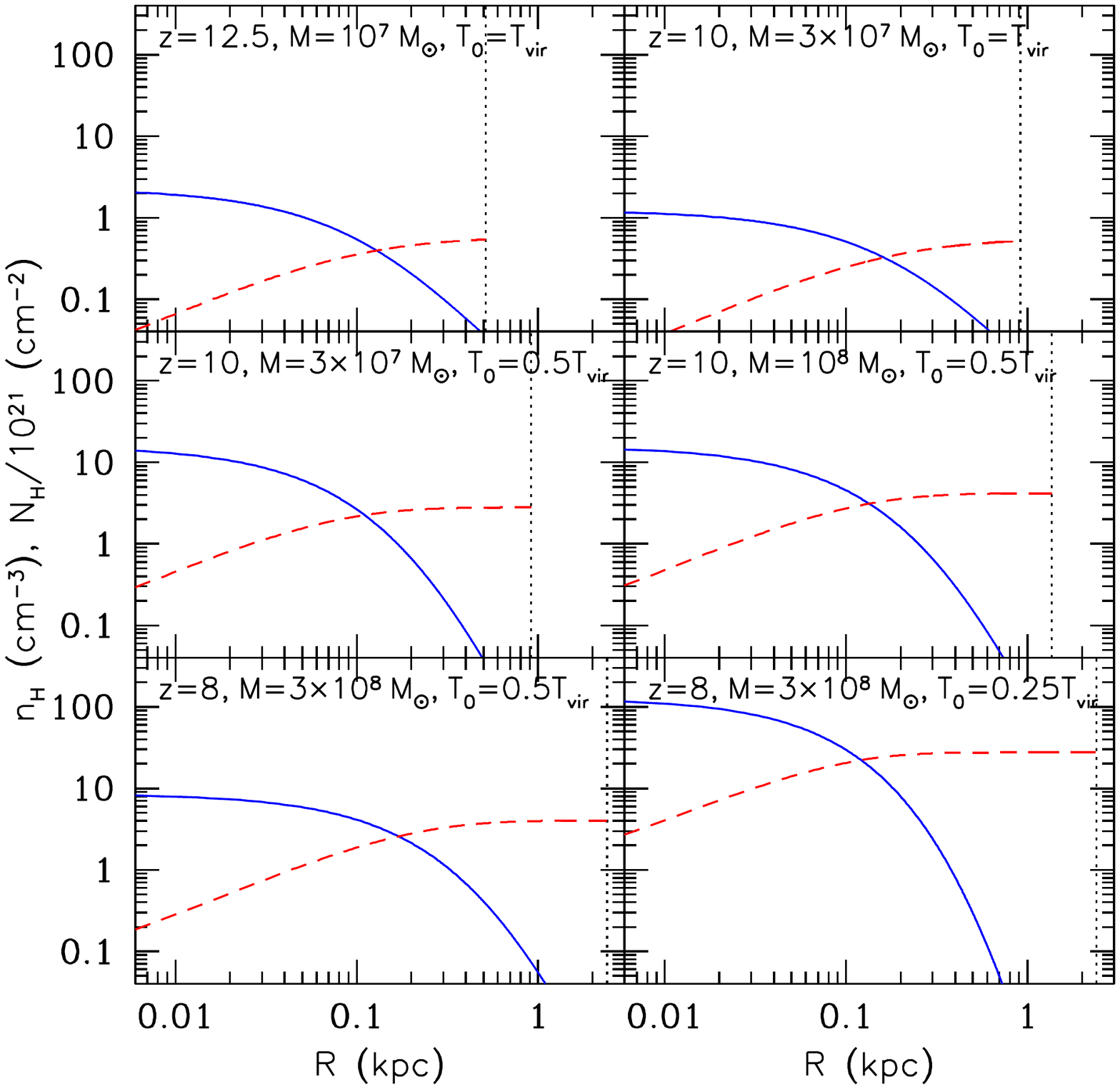}  
\caption{Hydrogen number density as a function of radius (blue solid
  lines) and column density within that radius (red dashed lines) for
  the galaxy models listed in Table~\ref{tab:galaxies} (the parameters of
  these models are also indicated in the panels). The vertical dotted
  lines denote the virial radii of the galaxies.
}
\label{fig:dens_prof}
\end{figure}

For $z=12.5$, we considered a galaxy of total mass $M=10^7\Msun$,
since the bulk of star formation at this epoch was probably taking
place in minihaloes with masses of this order. As typical
star-forming galaxies were becoming more massive with decreasing
redshift, we adopted $M=3\times 10^7$ and $10^8\Msun$ for $z=10$, and
$M=3\times 10^8\Msun$ for $z=8$. In reality, even more massive haloes,
$M\gtrsim 10^9\Msun$, likely provided a significant contribution to
the global SFR at $z\lesssim 10$ (e.g. \citealt{xuetal16}), but we do not
consider such galaxies in the present study since our adopted
spherically symmetric gas density distribution (implying
unrealistically high central densities for reasonable gas temperatures
$\sim 10^4$~K) is probably not a good approximation in this case owing
to the likely formation of a central disc
(e.g. \citealt{rometal11,pawetal13}). 

The next parameter (see \S\ref{s:gas}) is the initial gas
temperature. Here, we adopted $T_0=\Tvir$ for the $M=10^7\Msun$
minihalo and two alternative values, $T_0=\Tvir$ and $T_0=0.5\Tvir$,
for the $M=3\times 10^7\Msun$ halo. In these cases, the initial gas
temperature is somewhat lower than $10^4$~K, the temperature at which atomic
cooling becomes efficient, and such regime is the defining property
of minihaloes. By assuming $T_0=0.5\Tvir$, we wish to explore how our
results may change if the ISM has been able to significantly cool
down (and hence to concentrate stronger toward the halo's centre) upon
virialisation. For our more massive galaxies, we adopted
$T_0=0.5\Tvir$ and/or $T_0=0.25\Tvir$ (see Table~\ref{tab:galaxies}),
so that the initial gas temperature was always $\sim 10^4$~K. We did
not allow $T_0>2\times 10^4$~K, since atomic cooling is expected to
quickly bring the ISM temperature down to $\sim 10^4$~K upon
virialisation should the virial temperature (i.e. the initial
temperature of the shocked gas) be significantly higher. 
Fig.~\ref{fig:dens_prof} shows the hydrogen number density
radial profiles and the corresponding column densities for our galaxy
models. 

In reality, in those models with $T_0<10^4$~K, we calculated the gas
density profile by substituting $T_0$ (as quoted in
Table~\ref{tab:galaxies}) into equation~(\ref{eq:rhogf}) but set the
ISM temperature at $10^4$~K at the onset of X-ray irradiation. This
was done because: (i) {\sc Cloudy} is not well suited for
low-temperature ($T\lesssim 10^4$~K) regime and (ii) the ULX will
quickly (within $\sim 10^3$~years) heat up the ISM over the volume of
interest to $10^4$~K, since atomic cooling is inefficient at
lower temperatures. We also note that the virial temperature ($\Tvir$)
entering the above parametrisation in terms of $T_0/\Tvir$ was derived
assuming a neutral medium, i.e. $\mu=1.2$ in equation~(\ref{eq:tvir}),
even if the actual initial (coronal-equilibrium) state of the ISM (as
calculated by {\sc Cloudy}) was a significantly ionised one (as is the
case for our hottest model with $T_0\approx 1.9\times 10^4$~K).   

We next adopted 5 fiducial values of the luminosity of the ULX:
$\Lx=10^{39.5}$, $10^{40}$, $10^{40.5}$, $10^{41}$ and
$10^{41.5}$~erg~s$^{-1}$. Here, the luminosity is the bolometric one,
for a spectrum given by
\beq
\frac{dN_\gamma}{dE}\propto E^{-2.1}e^{-100~{\rm eV}/E-E/10~{\rm keV}},
\label{eq:spec}
\eeq
which is slightly different from the power-law spectrum with sharp
boundaries at 100~eV and 10~keV that we adopted for the analytical
estimates in \S\ref{s:estimates}. Such a modification is supposed to
render the numerical calculations better-behaved because of the absence
of discontinuities in the spectrum of incident radiation. 

For the spectrum given by equation~(\ref{eq:spec}), the fraction of
the bolometric luminosity contained in the 0.1--10~keV and 0.25--8~keV
bands (the latter was used in \citealt{sazkha17b,sazkha17c} for
construction of the X-ray luminosity function and collective X-ray
spectrum of HMXBs in the local Universe) is 0.87 and 0.69, respectively.

We assumed the ISM to consist purely of hydrogen and helium, although
we also ran a model with a non-negligible contribution of metals (at
the level of 0.1 of the solar value, see \S\ref{s:metal} below). In
both cases, molecular and grain physics were switched off by the
corresponding options of the code. We adopted a spherical geometry
with a density distribution given by equations~(\ref{eq:rhogf}) and
(\ref{eq:rhog0}) within $\Rvir$ (and no gas outside). The X-ray source
was located at the center of the gas cloud. 

The calculation starts with finding coronal equilibrium
in the gas with uniform temperature $T_0$. After that, the ULX is
switched on (at $t=0$) and allowed to shine at a constant luminosity
$\Lx$ for a long time (specifically, we ran the computations up to
$t=10^7$~yr), well in excess of realistic ULX lifetimes ($\sim
10^5$~yr). As a result of illumination by the central source,
the ionisation state, temperature and opacity of each shell evolve in
time, and the code follows this evolution with the output produced at
logarithmically spaced time steps. 

At every iteration, we save physical conditions (temperature, H and He
ionisation states etc.) across the whole computational domain along
with the integrated energy-dependent optical depths and spectra of
the radiation (the transmitted radiation of the central source and the
radiation emitted by the photoionised gas) as it
emerges at the outer boundary of the sphere. Comparing the properties of
this emergent radiation with the intrinsic luminosity and spectrum of
the central source, we manage to self-consistently track the evolution
of the photoabsorption of X-rays from the ULX in its host galaxy.

\section{Results}
\label{s:results}

\begin{figure}
\centering
\includegraphics[width=\columnwidth,viewport=20 180 560 720]{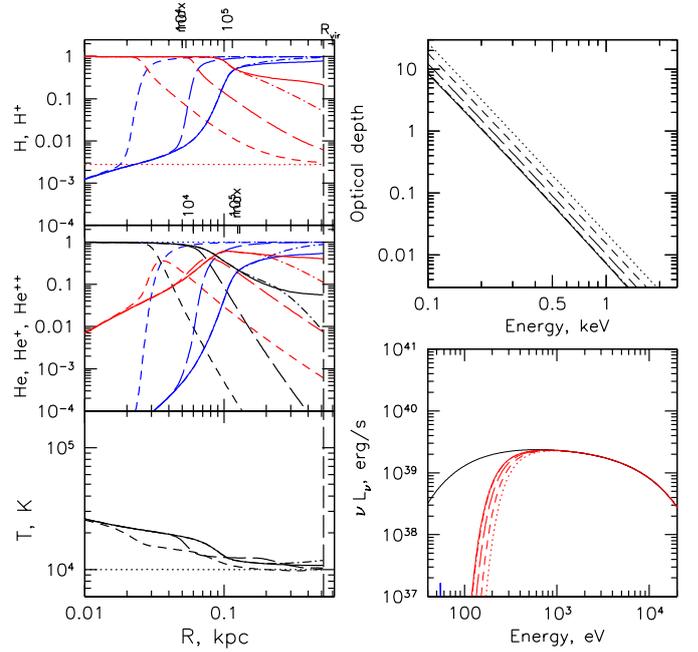} 
\caption{Results of computations for galaxy model 1 (see
  Table~\ref{tab:galaxies}) and ULX luminosity
  $\Lx=10^{40}$~erg~s$^{-1}$. {\sl Top left:} Radial profiles of the
  HI and HII relative fractions (blue and red, respectively) as a
  function of time: $t=0$ (dotted), $t=10^4$~yr (short-dashed),
  $t=10^5$~yr (long-dashed), $t=10^6$~yr (dash-dotted) and
  $t=10^7$~yr~(solid). The same line types are used in the other
  panels. Also shown (see the marks at the top of the panel) are the
  virial radius and the analytical estimates of the radius of the HII
  zone at $t=10^4$, $10^5$~yr and in photoionisation equilibrium,
  given by eqs.~(\ref{eq:rHIIpc}) and (\ref{eq:rHIImaxpc}),
  respectively. These were calculated for the $\nH$ at the
  centre of the galaxy. {\sl Middle left:} 
  The same for the HeI (blue), HeII (red) and HeIII (black) relative
  fractions. The ticks at the top of the panel show the analytical
  estimates for $\RHeIII$ at $t=10^4$, $10^5$~yr and $\RHeIIImax$ according to
  eqs.~(\ref{eq:rHeIIIpc}) and (\ref{eq:rHeIIImaxpc}),
  respectively. {\sl Bottom left:} Evolution of the gas temperature
  radial profile. {\sl Top right:}. Evolution of the photoionisation
  optical depth (as a function of energy). {\sl Bottom right:}
  Evolution of the transmitted X-ray spectrum (red lines). The blue
  lines (if present) show the spectrum
  of the radiation emitted by the photoionised gas. The black line
  shows the intrinsic ULX spectrum.
}
\label{fig:model1_lx40}
\end{figure}

\begin{figure}
\centering
\includegraphics[width=\columnwidth,viewport=20 180 560 720]{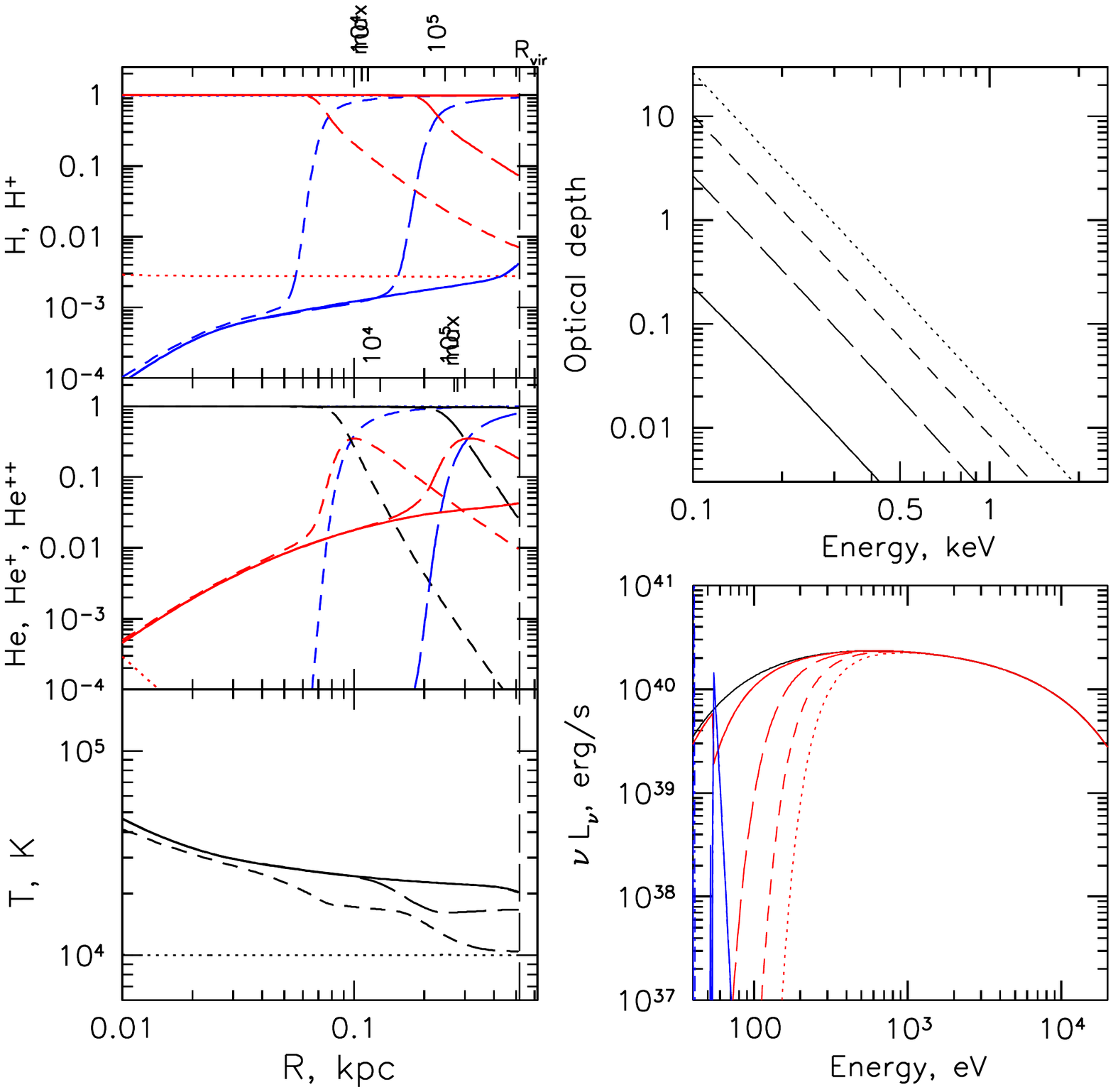}
\caption{As Fig.~\ref{fig:model1_lx40}, but for galaxy model
  1 (see Table~\ref{tab:galaxies}) and $\Lx=10^{41}$~erg~s$^{-1}$.
}
\label{fig:model1_lx41}
\end{figure}

\begin{figure}
\centering
\includegraphics[width=\columnwidth,viewport=20 180 560 720]{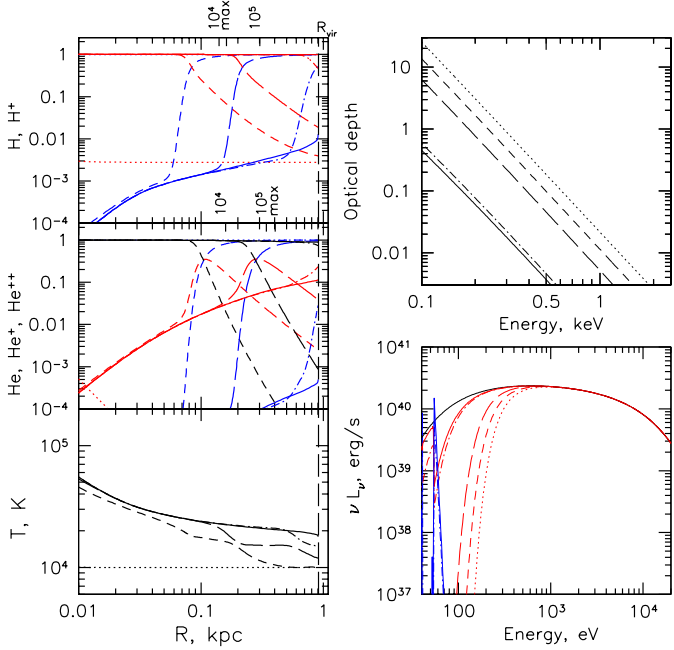}
\caption{As Fig.~\ref{fig:model1_lx40}, but for galaxy model
  2 (see Table~\ref{tab:galaxies}) and $\Lx=10^{41}$~erg~s$^{-1}$.
}
\label{fig:model2_lx41}
\end{figure}

\begin{figure}
\centering
\includegraphics[width=\columnwidth,viewport=20 180 560 720]{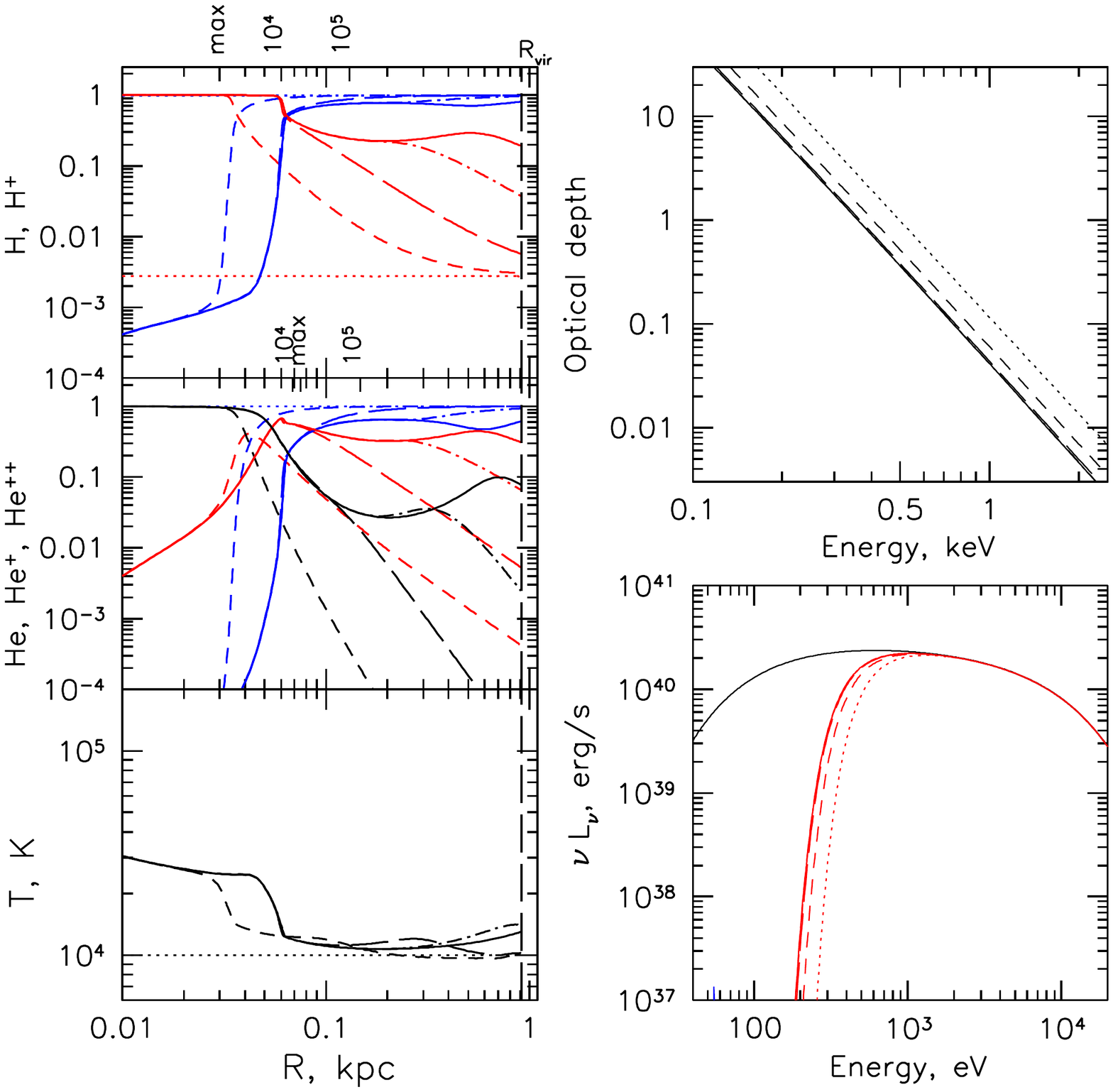} 
\caption{As Fig.~\ref{fig:model1_lx40}, but for galaxy model 3
  (see Table~\ref{tab:galaxies}) and $\Lx=10^{41}$~erg~s$^{-1}$.
}
\label{fig:model3_lx41}
\end{figure}

\begin{figure}
\centering
 \includegraphics[width=\columnwidth,viewport=20 180 560 720]{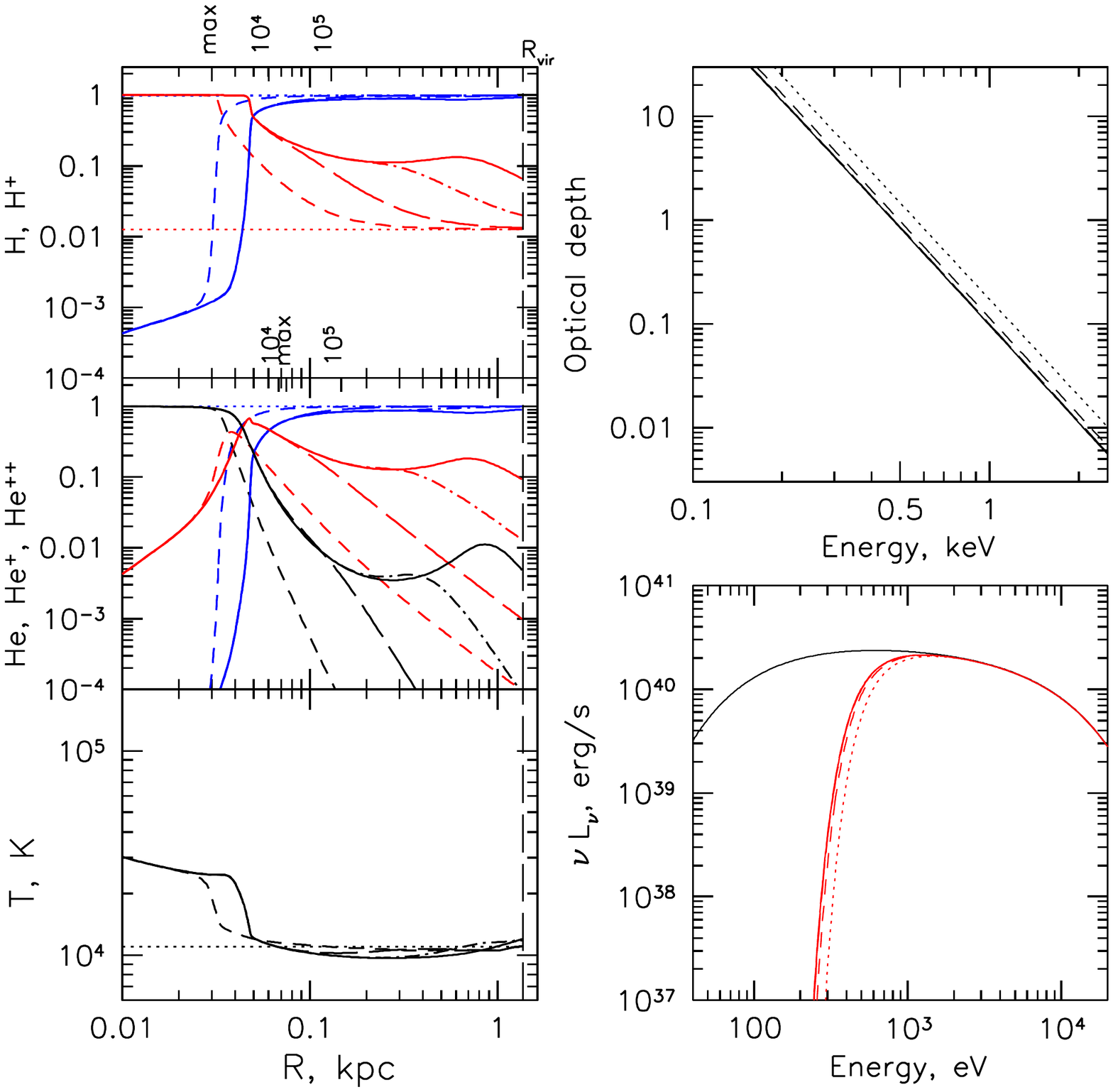} 
\caption{As Fig.~\ref{fig:model1_lx40}, but for galaxy model 4
  (see Table~\ref{tab:galaxies}) and $\Lx=10^{41}$~erg~s$^{-1}$.
}
\label{fig:model4_lx41}
\end{figure}

\begin{figure}
\centering
  \includegraphics[width=\columnwidth,viewport=20 180 560 720]{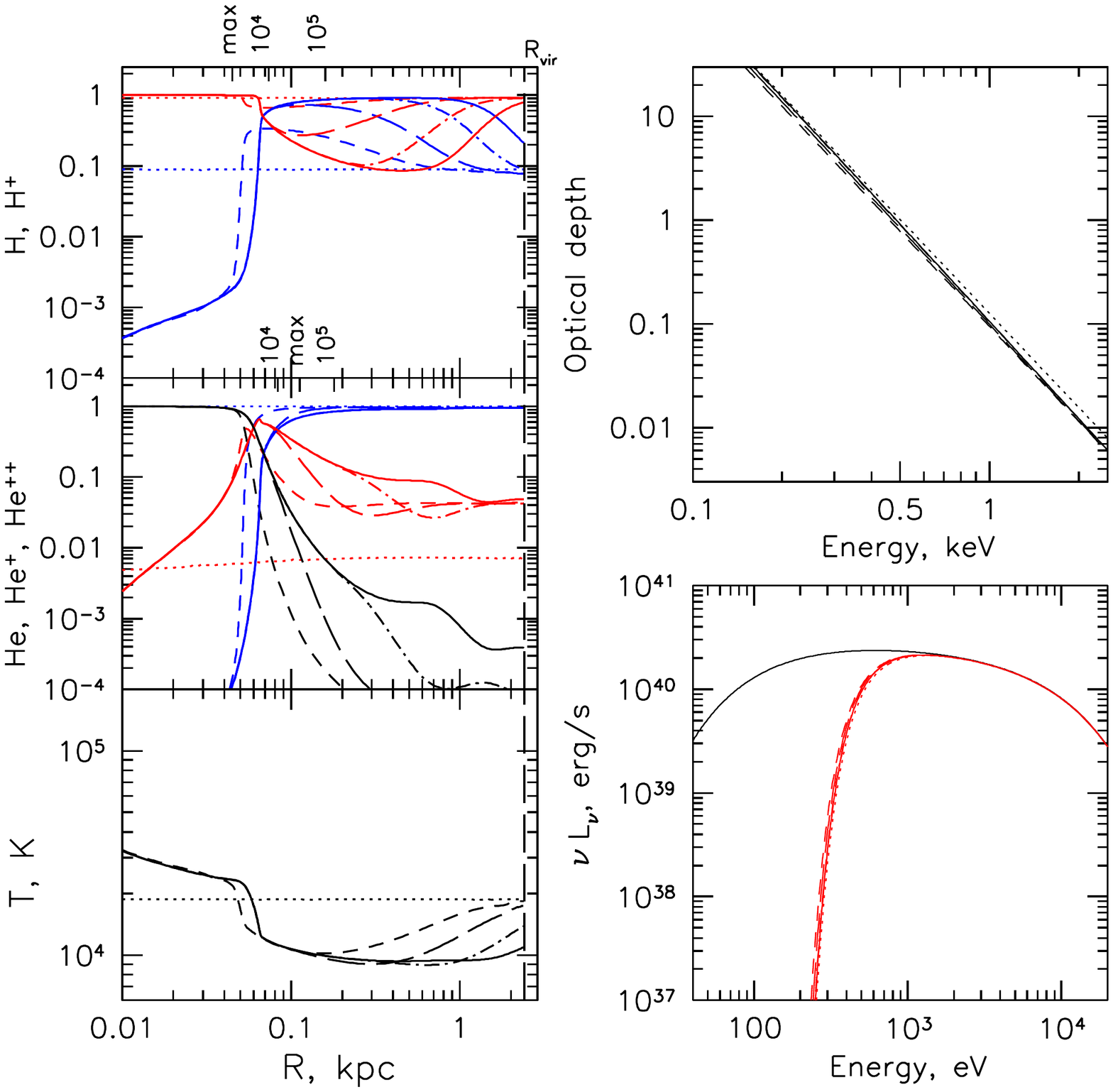}
\caption{As Fig.~\ref{fig:model1_lx40}, but for galaxy model 5
  (see Table~\ref{tab:galaxies}) and $\Lx=10^{41}$~erg~s$^{-1}$.
}
\label{fig:model5_lx41}
\end{figure}

\begin{figure}
\centering
  \includegraphics[width=\columnwidth,viewport=20 180 560 720]{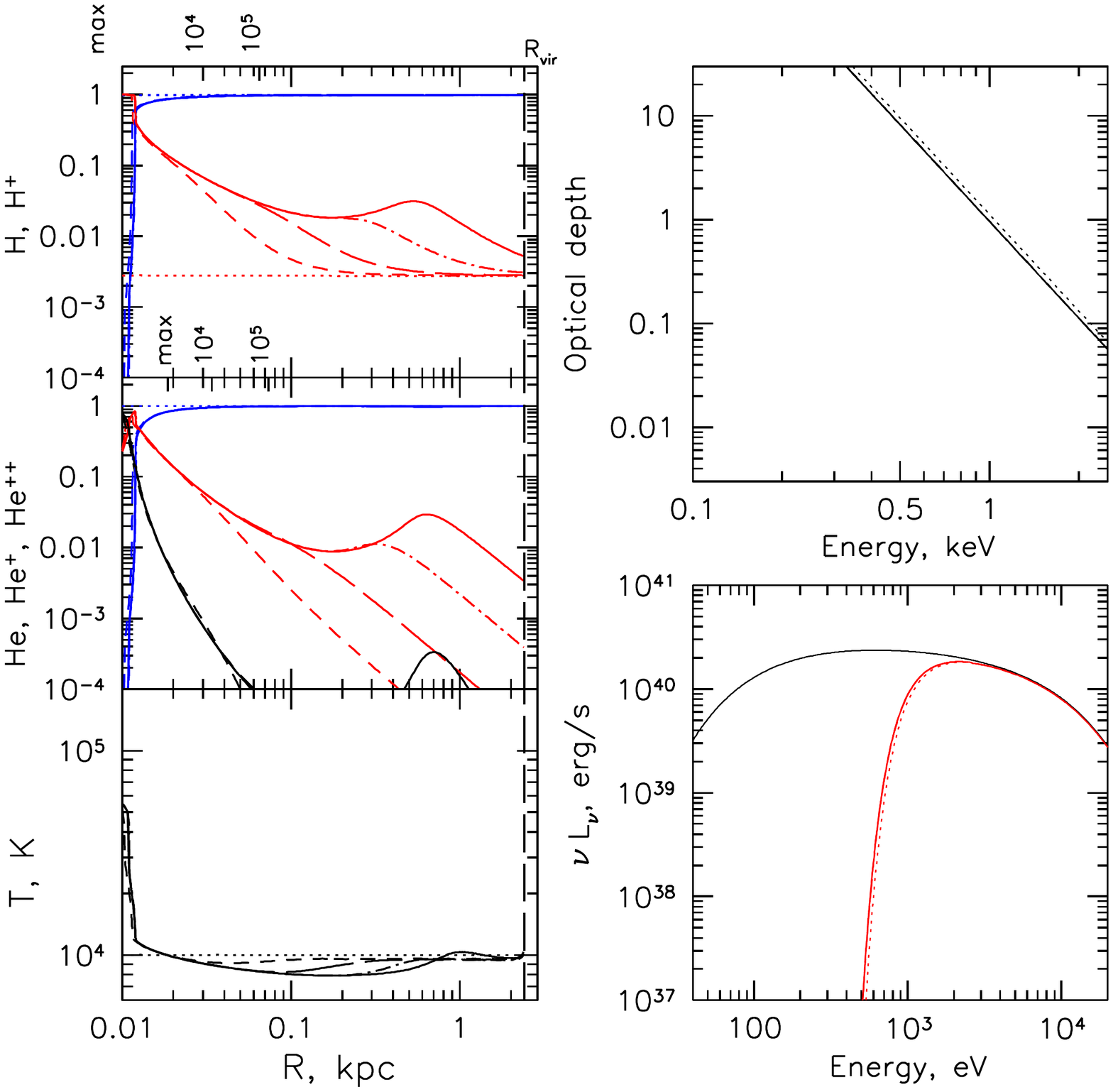}
  \caption{As Fig.~\ref{fig:model1_lx40}, but for galaxy model 6
  (see Table~\ref{tab:galaxies})  and $\Lx=10^{41}$~erg~s$^{-1}$.
}
\label{fig:model6_lx41}
\end{figure}

Figures~\ref{fig:model1_lx40}--\ref{fig:model6_lx41} show the results
of our numerical calculations for a subsample of the considered
galaxy/ULX models. Each figure shows the evolution of the radial
profiles of the HI, HII, HeI, HeII and HeIII relative fractions, gas
temperature and photoionisation optical depth, as well as the
evolution of the transmitted X-ray spectrum. Specifically, the radial
profiles and spectra are shown for elapsed times $t=0$, $10^4$,
$10^5$, $10^6$ and $10^7$ years.

In the panels for hydrogen and helium, we also show analytical
estimates of the radii of the HII and HeIII zones based on the
expressions presented in \S\ref{s:estimates}. To this end, we
substituted the halo's central hydrogen density (see the last column
in Table~\ref{tab:galaxies}) for $\nH$. We show  $\RHII$ and
$\RHeIII$ at $t=10^4$ and $10^5$~yr, even if either radius becomes
larger than the expected maximum size of the corresponding HeII or
HeIII zone ($\RHIImax$ and $\RHeIIImax$, respectively), i.e. even if
$\tHII$ or $\tHeIII$ given by equations~(\ref{eq:tHII}) and
(\ref{eq:tHeIII}) prove to be shorter than $10^4$ or $10^5$~yr.

As expected, the less massive the galaxy is, the stronger is the
impact of the ULX on the ISM. Consider first our least massive halo
($M=10^7\Msun$, $z=12.5$), i.e. model 1 (see
Table~\ref{tab:galaxies}). As seen from Fig.~\ref{fig:model1_lx40}, a
ULX with a relatively low luminosity of $10^{40}$~erg~s$^{-1}$ can
significantly (by a factor of $\sim 2.5$) reduce the optical depth of
the ISM within $t\sim 10^5$~yr. As the photoionisation
front propagates outwards, the affected ISM is heated up to $\sim
2\times 10^4$~K, which induces additional, collisional ionisation of
the gas (in particular, hydrogen) and reduces its recombination
rate. As a result, our analytical estimates (in \S\ref{s:estimates})
for the maximum size of the HII zone ($\RHIImax$), derived under the
assumption of photoionisation equilibrium at $T=10^4$~K, become
inadequate, whereas our estimates for the time-dependent radii
$\RHII(t)$ and $\RHeIII(t)$ as well as for the maximum size of the
HeIII zone ($\RHeIIImax$) prove to be fairly accurate (see
Figs.~\ref{fig:model1_lx40}--\ref{fig:model6_lx41}). In the case of a
more powerful ULX, with $\Lx=10^{41}$~yr, the X-ray ionisation/heating
effects become more dramatic: the optical depth drops by an order of
magnitude within $10^5$~yr, as most of the minihalo is ionised after
this time (see Fig.~\ref{fig:model1_lx41}).

Consider next a somewhat larger halo, as represented by our model 2
($M=3\times 10^7\Msun$, $z=10$). In this case, the virial temperature
is $9.9\times 10^3$~K, at which atomic cooling is already important,
so that this object may be considered a transitional case between
minihaloes and dwarf galaxies. To make allowance for the uncertainty
in the actual initial temperature and radial density distribution of
the gas, we consider two cases, $T_0=\Tvir$ and $T_0=0.5\Tvir$. In the
former case, as seen from Fig.~\ref{fig:model2_lx41}, the ISM is
ionised nearly as strongly as in the 
case of the $10^7\Msun$ minihalo. Specifically, a ULX with
$\Lx=10^{41}$~erg~s$^{-1}$ can reduce the total
photoionisation optical depth by a factor of $\sim 5$ after $t\sim
10^5$~yr. The ionisation effect becomes less dramatic in the
$T_0=0.5\Tvir$ case (see Fig.~\ref{fig:model3_lx41}), since the
central gas density is now an order of magnitude higher than for
$T_0=\Tvir$ ($\nH=15.4$ vs. 1.2~cm$^{-3}$) and it is more difficult
for the ULX to ionise and heat the central regions of the halo. 


For the yet more massive halo represented by our model 4 ($M=10^8\Msun$,
$z=10$), the ionisation effect is still noticeable: a ULX with
$\Lx=10^{41}$~erg~s$^{-1}$ can reduce the total photoionisation
optical depth by a factor of $\sim 2$ after $t\sim 10^5$~yr (see
Fig.~\ref{fig:model4_lx41}). However, our most massive galaxy
($M=3\times 10^8\Msun$, $z=8$) is almost unaffected by X-ray
irradiation (see Figs.~\ref{fig:model5_lx41} and \ref{fig:model6_lx41}),
regardless of the actual configuration of the gas (compare the results
for model 5 with $T_0=0.5\Tvir=1.9\times 10^4$~K and model 6 with
$T_0=0.25\Tvir=9.4\times 10^3$~K). In the latest (densest) case, the ionisation
front is unable to propagate further out than $\sim 10$~pc, although
X-ray photons of relatively high energy ($E\gtrsim 1$~keV) are able to
propagate to large distances from the ULX and eventually raise the
ionisation fraction (i.e. the relative fractions of HII and HeII) to
a few per cent over the bulk of the galaxy.

\subsection{Impact of metal enrichment}
\label{s:metal}

The ISM at $z\sim 10$ may have already been weakly (to less than a few
per cent of the solar metallicity, e.g. \citealt{paletal14}) enriched by
metals. We have therefore repeated the computations for some of our
models adding a $Z=0.1$ (with respect to the solar abundances)
fraction of heavy elements to the primordial mixture of hydrogen and helium.

\begin{figure}
\centering
\includegraphics[width=\columnwidth,viewport=20 180 560 720]{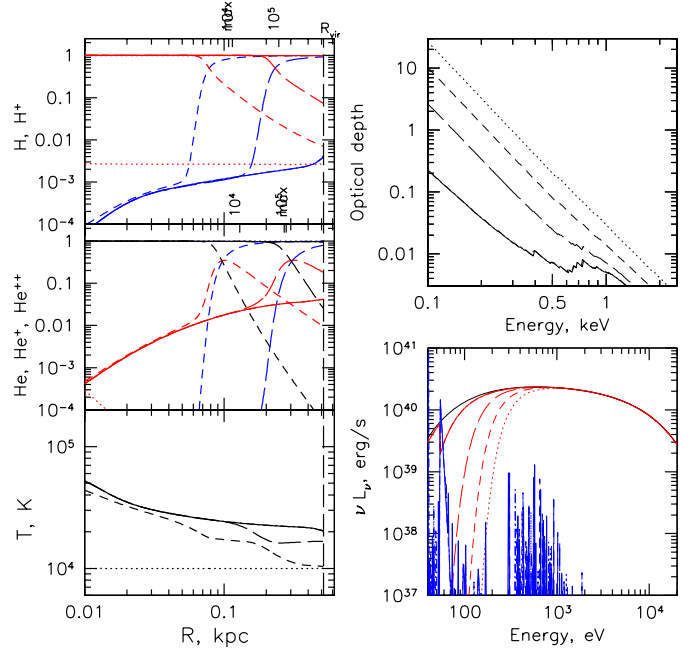}
  \caption{As Fig.~\ref{fig:model1_lx41} (galaxy model 1 and
    $\Lx=10^{41}$~erg~s$^{-1}$), but for non-zero metallicity $Z=0.1$. 
}
\label{fig:model1_lx41_z01}
\end{figure}

Figure~\ref{fig:model1_lx41_z01} shows an example of such a test. By
comparing these plots with those for the same galaxy/ULX model but
without metals (Fig.~\ref{fig:model1_lx41}), we see that the results
have remained almost unaltered. A difference is only noticeable at
energies above $\sim 500$~eV, where the metals (mainly oxygen and the
L-shell of iron) start to contribute significantly to the optical
depth of the ISM once the hydrogen and helium have been photoionised
by the ULX. However, since the ISM optical depth at these energies is
already very small by this moment, the overall difference in the
transmitted soft X-ray flux (see below) between the metal-poor and
metal-free cases is not significant.

\subsection{Ultimate effect on the transmitted X-ray flux}
\label{s:flux}

The bottom right panels of
Figs.~\ref{fig:model1_lx40}--\ref{fig:model6_lx41} show the X-ray
spectrum emergent from the galaxy at different times ($t=0$, $10^4$,
$10^5$, $10^6$ and $10^7$~yr). These spectra represent the ULX 
radiation transmitted through the ISM. The contribution of the
emission produced by the ISM itself (also shown in 
the figures) is negligible in the considered energy range ($\gtrsim
50$~eV), since the gas always remains colder that $\sim 7\times 10^4$~K.

As scattering on free electrons provides a negligible contribution to
the total optical depth through the gas, the emergent X-ray spectrum
is approximately equal to the incident spectrum [given by
  eq.~(\ref{eq:spec})] multiplied by $\exp[-\tau(E)]$, where
$\tau(E)$ is the photoabsorption optical depth, plotted in the top
right panels of Figs.~\ref{fig:model1_lx40}--\ref{fig:model6_lx41}.
For the problem of X-ray heating of the IGM in the early Universe, the
most interesting property of a population of X-ray sources
(e.g. HMXBs), apart from its cumulative luminosity, is the 
fraction of the intrinsic soft X-ray (below $\sim 1$~keV) luminosity
escaping from the host galaxies. Note that because the IGM remains
nearly neutral during the X-ray heating epoch, it is mainly the
soft X-ray luminosity rather than the total number of soft X-ray
photons that is of importance, since secondary ionisations caused by
fast photoelectrons play a major role. This is in contrast to X-ray
irradiation of the ISM within galaxies, where the gas eventually
becomes significantly ionised so that harder X-ray photons release
most of their energy as heat rather than in multiple ionisations. 

\begin{figure}
\centering
\includegraphics[width=\columnwidth,viewport=20 150 560 720]{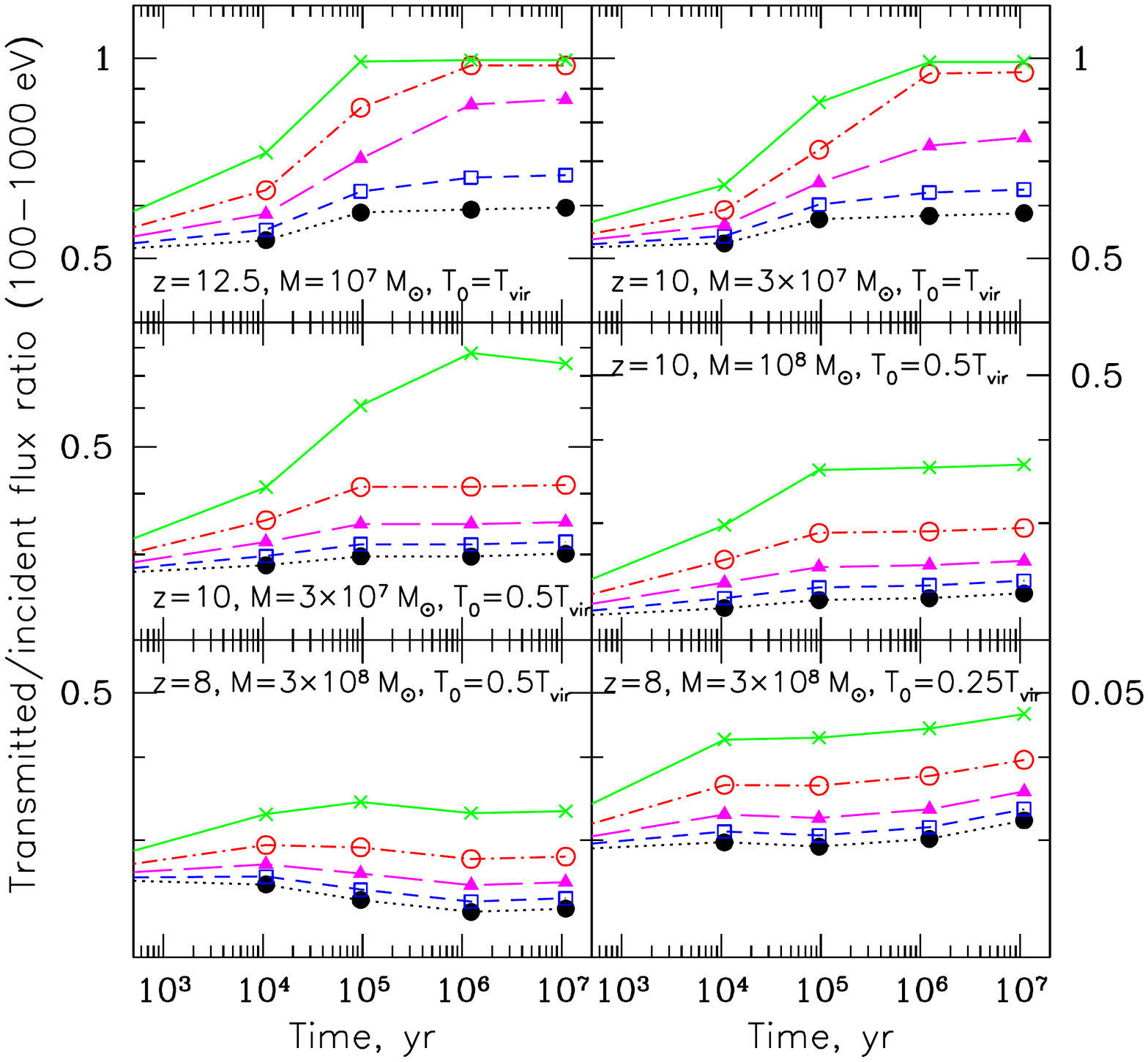} 
\caption{Fraction of the soft X-ray (0.1--1~keV) luminosity of the ULX 
  transmitted through the galaxy as a function of time elapsed since
  the switch-on of the source, for the different galaxy models (see
  Table~\ref{tab:galaxies} and the labels inside the panels), as a
  function of the bolometric luminosity of the ULX:
  $\Lx=10^{39.5}$~erg~s$^{-1}$ (black filled circles 
  connected by the dotted line), $\Lx=10^{40}$~erg~s$^{-1}$ (blue
  open squares connected by the short-dashed line),
  $\Lx=10^{40.5}$~erg~s$^{-1}$ (magenta filled triangles connected by
  the long-dashed line), $\Lx=10^{41}$~erg~s$^{-1}$ (red open circles
  connected by the dash-dotted line) and $\Lx=10^{41.5}$~erg~s$^{-1}$
  (green crosses connected by the solid line). Note the different
  vertical scales for different models.
}
\label{fig:flux_100_1000}
\end{figure}

Therefore, we determined for all of our galaxy/ULX models the ratio
$\Rs\equiv\Lsesc/\Ls$, where $\Ls$ is the intrinsic luminosity of the ULX in
the soft X-ray (0.1--1~keV) energy band (note that $\Ls/\Lx=0.48$ for
our adopted ULX spectrum) and $\Lsesc$ is the transmitted luminosity
in the same band. This ratio is shown in Fig.~\ref{fig:flux_100_1000}
for the sampled galaxy/ULX models and different elapsed times.

We see that for the $M=10^7\Msun$ minihalo, $\Rs\approx 50$\% at the
start of X-ray irradiation ($t=0$), and the ULX, depending
on its luminosity ($\Lx=10^{39.5}$--$10^{41.5}$~erg~s$^{-1}$),
gradually increases the escape fraction to $\sim
60$--100\% within $t=10^5$~yr. For our somewhat larger halo 
($M=3\times 10^7\Msun$) and $T_0=\Tvir=9.9\times 10^3$~K, the impact
of the ULX on $\Rs$ is somewhat weaker, and it weakeens further (but
remains significant) if the gas configuration is more compact
($T_0=0.5\Tvir$). For the more massive halo with $M=10^8\Msun$ (and
$T_0=0.5\Tvir$), a ULX with $\Lx=10^{41}$~erg~s$^{-1}$ can increase
$\Rs$ from $\approx 20$\% to $\approx 30$\% within $10^5$~yr. Finally,
for our most massive halo ($M=3\times 10^8\Msun$), X-ray irradiation
is unable to significantly affect the transmitted soft X-ray flux for
realistic ULX luminosities ($\Lx\lesssim {\rm
  a~few~}10^{41}$~erg~s$^{-1}$)\footnote{Note that for model 5, $\Rs$
  actually decreases with time for $\Lx\lesssim 10^{41}$~erg~s$^{-1}$,
  due to the ISM cooling to $\sim 10^4$~K from its initial (relatively high)
  temperature of $1.9\times 10^4$~K.}. 

\begin{figure}
\centering
\includegraphics[width=\columnwidth,viewport=20 150 560 720]{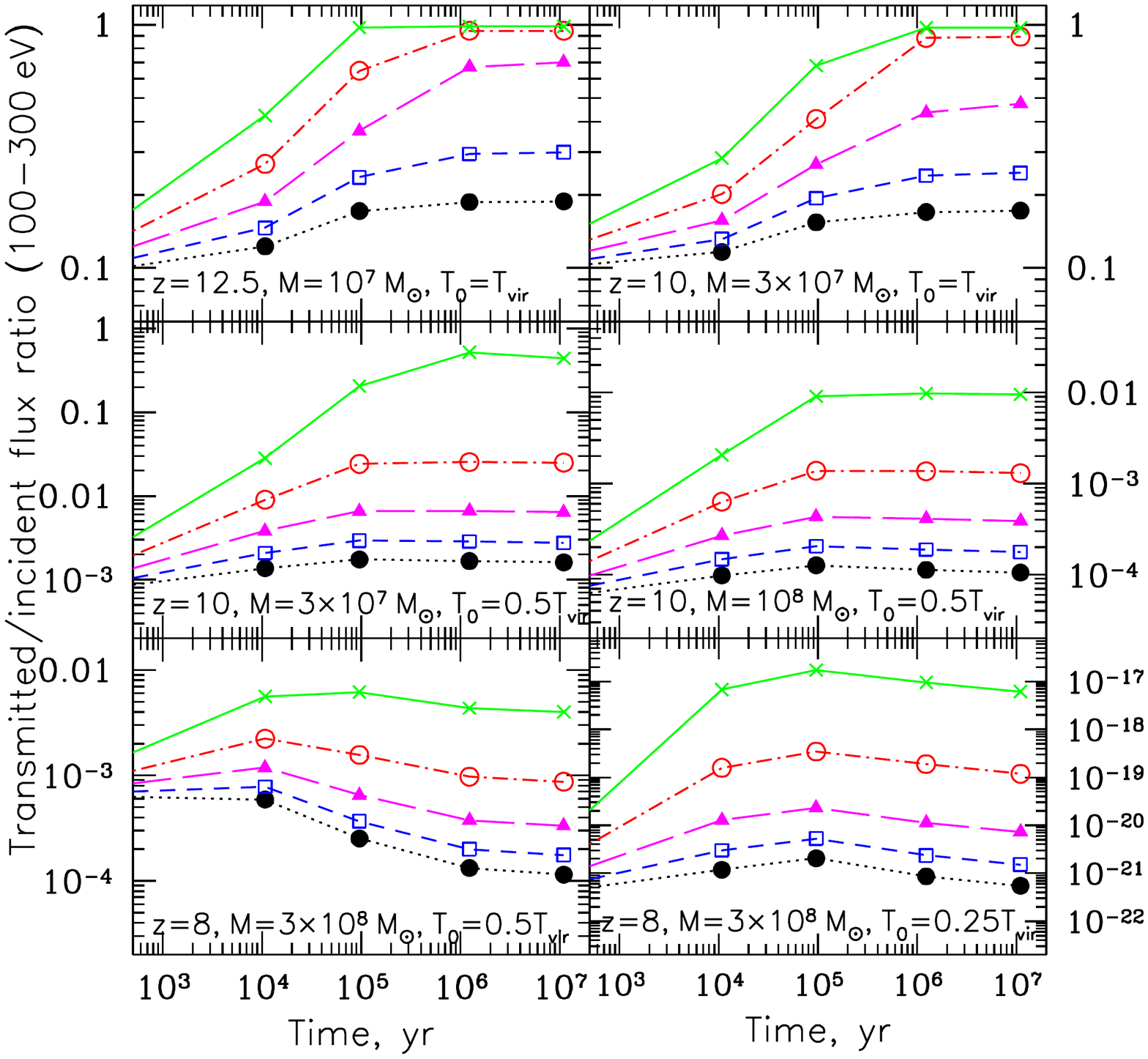} 
\caption{The same as Fig.~\ref{fig:flux_100_1000}, but for the
  supersoft (100--300~eV) energy band.
}
\label{fig:flux_100_300}
\end{figure}

In reality, X-ray heating is not expected to be homogeneous in the
early Universe, as the IGM in the close vicinity of strong X-ray
sources might be heated to somewhat higher temperatures than
elsewhere. Since the mean free path of X-ray photons increases 
dramatically with increasing photon energy, $\bar{\lambda} \sim
5[(1+z)/10]^{-3}(E/500~{\rm eV})^{3.2}$~proper~Mpc
(e.g. \citealt{sazkha17a}), it is only the softest photons that can 
play a role in such localised heating. It is thus
interesting to additionally evaluate the escape fraction for
the supersoft (100--300~eV) energy band: $\Rss\equiv\Lssesc/\Lss$,
where $\Lss$ and $\Lssesc$ are the intrinsic and transmitted ULX
luminosities in this band (note that $\Lss/\Lx=0.20$ for our adopted
ULX spectrum). The resulting dependences of $\Rss$ on $\Lx$ and $t$
for the different galaxy models are shown in Fig.~\ref{fig:flux_100_300}. 

As could be expected, the impact of the ULX on the escape fraction is
much more dramatic in the supersoft band. The $\Rss$ fraction
increases noticeably as a result of X-ray irradiation for all of our
galaxy models. It should be noted, however, that for the more massive
haloes ($M=10^8\Msun$ and $3\times 10^8\Msun$), the supersoft X-ray
escape fraction remains below 1\% even for the most luminous 
ULXs (in particular, $\Rss\approx 0$ for Model~6, due to the huge optical
depth). Therefore, soft X-rays from ULXs may cause a significant
extra heating of the IGM only near minihaloes and dwarf galaxies with
$M\lesssim {\rm a~few~}\times 10^7\Msun$. 

\section{Discussion}
\label{s:discuss}

This study is just a first step in the exploration of the potential
impact of ULXs/HLXs on the X-ray opacity of their host dwarf galaxies and
minihaloes in the early Universe. The 
galaxy models considered in this paper are admittedly over-simplified,
not least because direct observations of the first galaxies will not
become possible until the next generation of ground-based and
satellite-borne telescopes come into operation. 

The main assumption of our current model is that the ISM is in
single-temperature hydrostatic equilibrium in the dark matter
potential well. As demonstrated by detailed simulations
(e.g. \citealt{wisetal14,dasetal17,treetal17}), the actual gas
distribution is likely to be substantially inhomogeneous, so that gas
column densities may vary greatly from one viewing direction to
another for a given galaxy. This implies that if such a galaxy hosts a
luminous X-ray source, its radiation will preferentially escape
through low-density channels in the ISM, and this anisotropic escape
will be strengthened by feedback of the ULX on the ISM, since it is
easier for X-rays to ionise and heat a rarefied medium.

The X-ray mode of feedback discussed in this paper may
well work in concert with other types of back-reaction associated with star
formation, such as mechanical feedback (through winds and collimated
outflows) from X-ray binaries, including the ULXs responsible for the
X-ray feedback (e.g. \citealt{jussch12}), as well as radiative and mechanical
feedback from massive stars and supernovae. In principle, the
starburst episode associated with the appearance of a ULX could
rarefy, pre-ionise and heat the gas in the galaxy, helping the ULX to
ionise it (within the X-ray beam) more strongly afterwards. However,
the actual sequence of events is not well known. In particular, since
the majority of ULXs in the first galaxies were probably among the
highest-mass, and hence most rapidly evolving, stellar binaries, they
could inject their energy into the ISM before luminous supernovae
started to provide a substantial feedback. Moreover, the relative
importance of stellar and ULX feedback in the early Universe may have
varied greatly from one dwarf galaxy to another, since the latter is
dominated by the brightest, and hence very rare, ULXs (see
\S\ref{s:intro}). We refer the reader to the papers by
\cite{jussch12,artetal15} for a detailed discussion of these issues. 

Clearly, in order to more realistically evaluate the escape fraction
of soft X-rays from ULXs in the early Universe, it is necessary to
include the X-ray feedback described in this paper into detailed
simulations of star formation in the first galaxies
together with all other relevant physical processes. Moreover, the
present study is effectively limited to haloes with masses below $\sim
3\times 10^8\Msun$. More massive galaxies are likely to have central discs,
and the propagation of X-rays from ULXs located inside such discs
requires a separate investigation. 

As was explained in the introduction (\S\ref{s:intro}), only a
  tiny fraction ($\lesssim 1$\%) of galaxies in the early Universe
  are expected to host a ULX with $\Lx\gtrsim
  10^{40}$~erg~s$^{-1}$ at a given moment. However, assuming that such
  X-ray sources remain active for a time ($\tx\sim 10^5$~yr) that is
  much shorter than the duration of the corresponding starburst episode,
  $\tsb\sim 10^7$--$10^8$~yr, there is nevertheless a significant
  probability that a typical galaxy at $z\sim 10$ will host one or
  even a few bright ULXs over the whole history of its star formation
  activity. It is unlikely though that several ULXs appearing in the
  same galaxy will lead to a larger fraction of their X-ray emission
  escaping into the IGM, since their ionisation bubbles will be
  largely independent of each other (due to random directions of the
  X-ray beams and the time span between the bubbles being
  much longer than the characteristic recombination time of the ISM). 

In this connection, it is also worth noting that a typical
  galaxy in the early Universe would exhibit only relatively
  low-luminosity HMXBs at a given time. Indeed, given the HMXB LF
  measured in the local Universe \citep{minetal12,sazkha17b} and 
  assuming (perhaps optimistically) that the specific occurence rate
  of HMXBs was a factor of $\sim 10$ higher at $z\sim 10$, a typical
  galaxy at that epoch, with a SFR $\lesssim 0.01\Msun$~yr${-1}$, is
  not expected to have HMXBs with $\Lx>{\rm a~few}\times
  10^{37}$~erg~s$^{-1}$ at a given instant. The combined luminosity of
  such sources is therefore less than $10^{38}$~erg~s$^{-1}$. Based on
  the results of the present study, it is clear that such
  low-luminosity HMXBs should have virtually no impact on the ISM of
  their host galaxy.

It is also worth noting that although the present study was devoted to
the particular case of ULXs, its results may
also be interesting in considering other types of X-ray
sources that could heat the IGM in the early stages of cosmic
re-ionisation, such as growing supermassive black holes
(i.e. microquasars, see
e.g. \citealt{madetal04,ricost04,wyiloe04,thozar08}). Since in this
case more luminous and more prolonged (compared to ULXs) 
episodes of X-ray activity are possible, one may get an idea of
the expected impact on the ISM (and hence on the transmitted soft
X-ray flux fraction) by looking at the results of our computations for the
maximum considered values of the X-ray source parameters, namely
$\Lx\sim {\rm a~few~}10^{41}$~erg~s$^{-1}$ and $t\sim
10^6$--$10^7$~yr (see Figs.~\ref{fig:flux_100_1000} and
\ref{fig:flux_100_300}).

\section{Summary}
\label{s:summary}

If the analogy with the local Universe is correct, then ULXs with
(isotropic-equivalent) luminosity $\Lx\gtrsim 10^{40}$~erg~s$^{-1}$
are expected to provide the bulk of the X-ray emission associated with
star formation in the early Universe. We have evaluated the
photoionisation effect of such strong individual X-ray sources on the
ISM of their host galaxies.

We found that ULXs with $\Lx\gtrsim 10^{40}$--$10^{41}$~erg~s$^{-1}$
(0.1--10~keV) can significantly ionise the ISM (in particular,
helium), and thus reduce its opacity to soft X-rays, in minihaloes and
dwarf galaxies with total masses $M\sim 10^7$--$10^8\Msun$ at
$z\gtrsim 10$ provided that the X-ray source remains active for a
$\tx\sim 10^5$~yr, which, according to recent population synthesis 
studies of HMXBs, is not an unreasonably long time for a supercritical
accretion phase onto a neutron star or a stellar-mass black hole. As a
result, the fraction of the soft X-ray (0.1--1~keV) luminosity
escaping from a galaxy hosting a bright ULX can increase from $\sim
20$--50\% to $\sim 30$--80\% (depending on $M$ and $\Lx$) over the ULX
lifetime. This implies that ULXs (and hence HMXBs as a whole) can
induce a stronger heating of the IGM in the early Universe
compared to estimates neglecting X-ray feedback. 

ULX feedback is even more efficient in the supersoft X-ray energy
range (100--300~eV), with the escape fraction in this band increasing
several times as a result of X-ray irradiation of the ISM by a
ULX. This may allow ULXs located in minihaloes and dwarf galaxies at
$z\gtrsim 10$ to significantly heat the IGM in their vicinity. 

On the other hand, we have shown that larger galaxies with $M\gtrsim
3\times 10^8\Msun$ could not be significantly ionised even by  
the brightest ULXs in the early Universe. Since such galaxies probably
started to dominate the global SFR at $z\lesssim 10$, the overall
escape fraction of soft X-rays from the HMXB population probably
remained low, $\lesssim 30$\%, at these epochs.
 
\section*{Acknowledgments}

The research was supported by the Russian Science Foundation (grant
14-12-01315).
  

\bsp	
\label{lastpage}
\end{document}

%% file: galaxies_table.tex
\begin{table}
  \caption{Galaxy models
    \label{tab:galaxies}
  }
  \begin{tabular}{|c|c|c|c|c|c|c|c|}
    \hline
    \multicolumn{1}{|c|}{No.} &
    \multicolumn{1}{c|}{$z$} &
    \multicolumn{1}{c|}{$M$} &
    \multicolumn{1}{c|}{$T_0/\Tvir$} &
    \multicolumn{1}{c|}{$T_0$} &
    \multicolumn{1}{c|}{$\Rvir$} &
    \multicolumn{1}{c|}{$\nH (r=0)$} \\
    \multicolumn{1}{|c|}{} &
    \multicolumn{1}{c|}{} &
    \multicolumn{1}{c|}{($\Msun$)} &
    \multicolumn{1}{c|}{} &
    \multicolumn{1}{c|}{(K)} &
    \multicolumn{1}{c|}{(kpc)} &
    \multicolumn{1}{c|}{(cm$^{-3}$)} \\
    \hline
  1 & 12.5 & $10^7$              & 1      & $5.8\times 10^3$ & 0.52 & 2.3 \\
  2 & 10    & $3\times 10^7$ & 1      & $9.9\times 10^3$ & 0.91 & 1.2 \\
  3 & 10    & $3\times 10^7$ & 0.5   & $4.9\times 10^3$ & 0.91 & 15.4 \\
  4 & 10    & $10^8$              & 0.5   & $1.1\times 10^4$ & 1.36 & 15.4 \\
  5 &   8    & $3\times 10^8$ & 0.5   & $1.9\times 10^4$ & 2.40 & 8.5 \\
  6 &   8    & $3\times 10^8$ & 0.25 & $9.4\times 10^3$ & 2.40 & $1.25\times 10^2$ \\
  \hline
\end{tabular}

\end{table}